\documentclass[sn-mathphys-num]{sn-jnl}

\usepackage{graphicx}%
\usepackage{multirow}%
\usepackage{amsmath,amssymb,amsfonts,stmaryrd,mathtools}%
\usepackage{amsthm}%
\usepackage{mathrsfs}%
\usepackage[title]{appendix}%
\usepackage{xcolor}%
\usepackage{textcomp}%
\usepackage{manyfoot}%
\usepackage{booktabs}%
\usepackage{algorithm}%
\usepackage{algorithmicx}%
\usepackage{algpseudocode}%
\usepackage{listings}%
\usepackage{hyperref,bm}
\newcommand\norm[1]{\left\lVert#1\right\rVert}

\usepackage{subcaption}

\newtheorem{remark}{Remark}%

\begin{document}

\title[]{Parameterized crack modelling based on a localized non-intrusive reduced basis method}


\author*{\fnm{Margarita} \sur{Chasapi}}\email{margarita.chasapi@rwth-aachen.de}



\affil*{\orgdiv{Institute of Structural Analysis and Dynamics}, \orgname{RWTH Aachen University}, \city{Aachen}, \country{Germany}} 




\abstract{This contribution presents a model order reduction strategy for fast parametric modelling of problems with cracks formulated on spline discretizations. In the context of damage detection, parametric reduced order models (ROMs) are well suited for fast computations by establishing an efficient offline/online split of the simulation process. The problems of interest focus on geometric parameters that describe the crack configuration and may pose challenges to constructing efficient ROMs. This work proposes a framework based on non-intrusive reduced basis methods and a localization strategy tailored to parametric problems with moving discontinuities. The combined benefits of non-intrusive ROMs and localization enable accurate and efficient reduction with low online cost. We demonstrate the applicability of the ROM approach with benchmark tests on linear elastic problems discretized with splines and the extended isogeometric method (XIGA) for crack modelling. The results we obtain show the accuracy and real-time efficiency of the constructed reduced order models.}




\maketitle

\section{Introduction}\label{sec1}
In the past years, the development of model order reduction techniques and real-time efficient simulation tools has attracted a lot of attention in the computational mechanics community. In many applications, the solution of parameterized problems governed by partial differential equations (PDEs) requires multiple evaluations for different physical or geometrical configurations and the computational cost can become prohibitive when it comes to real-world engineering applications. In the context of structural monitoring, a typical challenge arises in the detection and prediction of damages: multiple simulations are often required to represent different damage configurations, such as cracks, incurring additional computational costs. The ability to rapidly track the current structural state requires tailored strategies that are both efficient and accurate. 

In the context of model order reduction techniques, reduced basis methods have been introduced in the past for PDEs. Originally, these rely on finite element approximations as high-fidelity problems and enable an efficient offline/online split, while a posteriori error estimators certify the accuracy of the constructed reduced basis approximation. The interested reader is referred to \cite{Hesthaven2016, QMN_RBspringer} for a detailed overview. Recently, such ROMs have been successfully adopted in the context of structural health monitoring to reduce the computational effort of constructing large data sets offline and repeatedly solving the underlying PDE in view of damage detection and localization in a real-time setting. To this end, ROMs were constructed based on the reduced basis method in \cite{Bigoni2020} by employing the acoustic-elastic wave equation and a Proper Orthogonal Decomposition (POD)-Galerkin approach. Moreover, parametric ROMs were constructed in \cite{Torzoni2022} based on a coupled thermo-mechanical model with parameterized loading and thermal conditions as well as in \cite{Torzoni2024} based on an elasto-dynamic model, while structural damage was represented by stiffness reduction. The above works employ finite element discretizations to construct ROMs and consider mainly physical parameters representing variations in the structural behavior, while geometric parameterization has not been thorougly investigated yet in this context. 

Reduced basis approximations were proposed in \cite{Huynh2007} to efficiently compute stress intensity factors for parameterized problems on domains with cracks considering both physical and geometric variations that are relevant to fracture mechanics applications. Thereby, parameterized linear elastic problems were handled based on piecewise-affine transformations. The latter hinder the application to complex geometries and general parameterizations, where the differential operators depend on the problem parameters in a non-affine manner. In general, non-affine problems with geometric parameters relevant to damage and fracture require further treatment to resolve the non-affine parametric dependence, such as efficient hyper-reduction methods \cite{Barrault2004,Chaturantabut2010,Negri2015}. In \cite{Agathos2020, Agathos2022} projection-based reduced order models were proposed based on the POD and mesh morphing for linear elastic cracked solids and shells. This approach exploits the extended finite element method \cite{Sukumar2000} to model the crack in combination with hyper-reduction techniques. The latter allow the efficient treatment of non-affine problems, although they are in principle intrusive and may require several modifications to high-fidelity solvers. Their combination with mesh morphing sucessfully tackles parametric crack problems formulated on finite element discretizations. Alternatively, enriching the reduced basis based on the algebraic approach in \cite{Agathos2024} allows efficient reduction for up to medium-scale problems. However, the mesh-independent treatment of general discretizations such as in higher-order and isogeometric methods still remains an open issue.

On another note, the introduction of Isogeometric Analysis (IGA) \cite{Hughes2005} has enabled the integration of geometric design with finite element analysis based on a common representation for both geometric modelling and numerical analysis. Computer Aided Design (CAD) commonly relies on spline basis functions, such as a B-splines or non-uniform rational B-splines (NURBS), for geometric representation. The main idea behind IGA is to adopt the same spline representation for the numerical solution within the finite element analysis. A detailed overview of the method can be found in \cite{Cottrell2009}. In the context of fracture, IGA is highly beneficial due to the higher continuity of splines and accuracy on a per-degree-of-freedom basis. Since its introduction it has been sucessfully employed in fracture mechanics, for instance, via the extended finite element method \cite{Sukumar2000, Luycker2011} and phase-field modelling \cite{Gomez2008, Borden2014, Ambati2015,Ambati2016}. 

In the past, there have been several works combining IGA with reduced basis methods \cite{Manzoni2015, Devaud2017, Salmoiraghi2016}. The interested reader is also referred to recent works on ROMs for multi-patch geometries \cite{Maquart2020, Chasapi2024c} and trimmed discretizations \cite{Chasapi2023, Chasapi2024}. These works rely on projection-based ROMs based on the POD and hyper-reduction techniques such as the Discrete Empirical Interpolation Method (DEIM) \cite{Chaturantabut2010, Negri2015} to resolve non-affine problems. This contribution falls within the context of non-intrusive reduced basis methods, while the construction of the reduced basis relies on the POD. Such non-intrusive ROMs facilitate learning from data \cite{Ghattas2021} and have been sucessfully applied, among other fields, in solid and structural mechanics \cite{Cicci2022, Tannous2025}. We further refer to interpolation-based ROMs exploiting the POD and their application to a wide range of problems \cite{Xiao2017,Hijazi2020,Georgaka2020, Garotta2020}. In particular when constructing a parametric ROM for crack detection-relevant problems, snapshot solutions exhibit moving discontinuities for different values of the parameters related to the crack configuration and the resulting solution manifold may be nonlinear. Thus, the construction of a global reduced basis leads to a very large number of basis functions and becomes inefficient. Local ROMs have been introduced in the past to overcome the above challenges. These are based on clustering of solution snapshots as first proposed in \cite{Amsallem2012}, further extended for hyper-reduction in \cite{Peherstorfer2014} and challenging applications in \cite{Pagani2018, Hess2019}. Adaptive local reduced bases have been also a subject of investigation in \cite{Haasdonk2011, Maday2013}. In the present work, a localization strategy is employed based on clustering of solution snapshots and learning multiple local reduced basis approximations in a non-intrusive manner.
This approach bears connections to local reduced bases introduced in \cite{Amsallem2012} and non-intrusive ROMs for nonlinear PDEs in \cite{Hesthaven2018, Geelen2022}, while it draws inspiration from our previous work on localized ROMs for trimmed spline discretizations \cite{Chasapi2023}. The reduction strategy is illustrated on numerical experiments employing the extended isogeometric method to model the crack. Note that other discretization methods are in principle also possible to model the discontinuity due to the crack. The main highlights of the proposed strategy are summarized as:
\begin{itemize}
\item It is non-intrusive and exploits data from the solution of high-fidelity problems that are already computed in the offline phase for the construction of the reduced basis.
\item It allows an efficient offline/online split for problems with geometric parameters that describe the crack in order to track variations of the crack configuration in an efficient and accurate manner. The non-intrusive ROM allows fast evaluations in the online phase for scenarios with non-affine parametric dependencies.
\item It enables efficient model reduction with localization based on clustering and classification techniques. The dimension of the local problems is small, which ensures the efficiency in the online phase.
\item It is suitable for a wide range of discretization techniques with respect to the underlying high-fidelity problem including higher-order and isogeometric methods.  
\end{itemize}

This contribution is structured as follows: Sec.~\ref{sec2} provides the necessary definitions related to the model problem of linear elasticity that we will consider thoughout this work. Sec.~\ref{sec3} presents a brief overview of B-spline basis functions and the parameterized solution discretization with XIGA. The localized non-intrusive reduced basis framework is elaborated in Sec.~\ref{sec4}. Several numerical experiments are carried out in Sec.~\ref{sec5}, while the main results and conclusions that can be drawn from this study are finally summarized in Sec.~\ref{sec6}.   

\section{Model problem}\label{sec2}
In the following, we briefly review the governing equations for linear elasticity that will serve as a model problem throughout this work. Let us consider a homogeneous, isotropic elastic body $\Omega \subset \mathbb{R}^d$, where $d$ denotes the dimension of the physical space.  The boundary of the domain is assumed to consist of two disjoint parts $\Gamma_D$ and $\Gamma_N$, representing the Dirichlet and Neumann boundary, accordingly. Therefore it holds $\overline{\Gamma}_D\cup\overline{\Gamma}_N = \partial{\Omega}$ and $\Gamma_D \cap \Gamma_N = \varnothing$. For simplicity, homogeneous Dirichlet boundary conditions are assumed throughout this work. The strong formulation of the problem reads: find $\boldsymbol{u}\in \left[ H^1_{0,\Gamma_D} \left( \Omega \right) \right]^d$ such that
\begin{equation}\label{elastic}
	\begin{cases}
        \begin{aligned}
		- \text{div} (\boldsymbol{\sigma}(\boldsymbol{u})) &= \tilde{\bm{f}} \qquad \qquad \qquad \qquad \qquad \ &&\text{in} \ \Omega \\
		\qquad \ \ \boldsymbol{\sigma}(\boldsymbol{u}) &= \mathbb{C} : \boldsymbol{\varepsilon}(\boldsymbol{u}) \ \ \quad \quad \quad \quad \ \ \ \ \ \ \ \   &&\text{in} \ \Omega \\
		\qquad \ \ \boldsymbol{\varepsilon}(\boldsymbol{u}) &= \displaystyle \frac{1}{2}(\nabla{\boldsymbol{u}} + (\nabla{\boldsymbol{u}})^T) \ \qquad \ \ \ \ \ &&\text{in} \ \Omega \\
		\qquad \ \  \boldsymbol{u} &= \boldsymbol{0} \qquad \qquad \qquad \qquad \qquad \ \ &&\text{on} \ \Gamma_D \\
		\qquad \ \ \boldsymbol{\sigma}(\boldsymbol{u})\cdot{\boldsymbol{n}} &= \tilde{\bm{g}} \qquad \qquad \qquad \qquad \qquad \ \  &&\text{on} \ \Gamma_{N},
		\end{aligned}
	\end{cases}
\end{equation}
where $\boldsymbol{\sigma}$ is a stress tensor, $\boldsymbol{\varepsilon}$ is a small strain tensor, $\tilde{\bm{f}}$ is the body force vector, $\boldsymbol{u}$ is the unknown displacement field, $\mathbb{C}$ is the material tensor double-contracted with the strain tensor $\boldsymbol{\varepsilon}$, $\boldsymbol{n}$ is the outward normal to the boundary and $\tilde{\bm{g}}$ is a prescribed traction on the Neumann boundary.  

The discrete weak formulation of the problem in Eq.~\eqref{elastic} can be expressed as: find $\boldsymbol{u}_h \in V_h$ such that 
\begin{equation}\label{weak}
	a(\boldsymbol{u}_h,\boldsymbol{v}_h)={f}(\boldsymbol{v}_h), \;\;\;\; \forall\boldsymbol{v}_h\in V_h,
\end{equation}
where $V_h \subset \left[ H^1_{0,\Gamma_D} \left( \Omega \right) \right]^d$ is a finite-dimensional vector subspace. The respective bilinear form $a(\cdot,\cdot)$ and the linear functional $f(\cdot)$ are given as:
\begin{equation}\label{weak2}
	a(\boldsymbol{u}_h,\boldsymbol{v}_h)=
	\int_\Omega
	\boldsymbol{\sigma}(\boldsymbol{u}_h):
	\boldsymbol{\varepsilon}(\boldsymbol{v}_h)\;
	d\Omega, \quad 
	{f}(\boldsymbol{v}_h)=
	\int_\Omega
	\tilde{\bm{f}}\;
	\boldsymbol{v}_h\;
	d\Omega + 	\int_{\Gamma_{N}}
	\tilde{\bm{g}}\;
	\boldsymbol{v}_h\;
	d\Gamma.
\end{equation}

\section{Extended isogeometric analysis for parameterized problems}\label{sec3}
The problems of interest throughout this work involve modelling discontinuities such as cracks in a parameterized setting, where multiple solutions of the problem in Eq.~\eqref{weak} are required for different values of the parameters. In the following, we will consider XIGA to model discontinuous and singular fields by locally augmenting the solution field with enrichment functions. Note that other discretization techniques are also possible to model the discontinuity due to the crack. An elaborate overview of extended finite element and isogeometric methods is provided in \cite{Sukumar2000, Nguyen2008, Luycker2011, Noel2022}.

\subsection{B-spline basis functions}\label{sec3_1}
In what follows, we will consider B-splines for the geometry description and solution discretization of our model problem. This section provides a basic review of B-splines, while a more detailed exposition on isogeometric analysis is given in \cite{Hughes2005, Cottrell2009}. For this purpose, we introduce a \emph{knot vector} $\Xi = \{\xi_1,\dots,\xi_{n+p+1}\}$ with each knot $\xi_i \in [0,1]^{{d}}$ and a univariate B-spline basis function $b^j_{i_j,p_j}$. Note that the integers $p$ and $n$ represent the degree and the number of basis functions, accordingly, while $p_j$ is the degree and $i_j$ the index of the basis function in the $j$-th parametric direction. The B-spline basis can be then defined based on the tensor product of univariate B-splines as:
\begin{equation}\label{spline}
\mathcal{B}(\boldsymbol{\xi}) = \prod_{j=1}^{{d}} b_{i_j,p_j}^j(\xi^j),
\end{equation}
where the dimension of the parametric space $d$ is assumed to be the same as the dimension of the physical space. Furthermore, $\boldsymbol{\xi}$ contains the parametric coordinates $\xi^j$ in the $j$-th parametric direction for $j=1,\dots,d$. In what follows the polynomial degree is  assumed to be equal in all parametric directions. We remark that the B-spline basis is $C^{p-k}$-continuous at knots, with $k$ being the multiplicity of the knot, while it is $C^{\infty}$ elsewhere. For a more elaborate overview of B-splines and NURBS, we encourage the interested reader to refer to \cite{Piegl1995}. Given the definition of B-splines in Eq.~\eqref{spline}, the geometric map ${\bf{F}}$ that describes the geometry of the domain $\Omega$ can be expressed as:
\begin{equation}\label{splinemap}
	{\bf{F}}(\boldsymbol{\xi}) = \sum_{\bf{i}} \mathcal{B}_{{\bf{i}},p}(\boldsymbol{\xi})\bf{P_i},
\end{equation}
where the vector $\bf{P_i}$ contains the coordinates of the control points. 

\subsection{Crack geometry representation}\label{sec3_2a}
In this work, the geometry of the crack is represented by Level Set Functions (LSFs). Although the XIGA approach is not restricted to a particular geometry representation, 
LSFs allows to handle straight and curved interfaces for fracture and multi-material problems \cite{Noel2022}. In the context of crack growth, the treatment of intersecting and branching cracks is in principle possible with tailored level set methods \cite{Moes2011}. 

The LSFs allow an implicit representation of the crack geometry. Let us first introduce a parameter vector $\bm{\mu} \in \mathcal{P} \subset \mathbb{R}^P$, where $\mathcal{P}$ is the parameter space and $P$ its dimension. In what follows, we will consider geometric parameters $\bm{\mu}=[\bm{\mu}_{cr} \ \bm{\mu}_{ct}]$ that describe the crack and crack tip, such as the position, length of the crack and location of the crack tip. Note that the domain $\Omega$ itself is assumed to be $\bm{\mu}$-independent without loss of generality. Moreover, it is assumed that the crack is traction free and no Dirichlet boundary conditions are imposed on the crack itself. In the numerical experiments of this work we consider straight-edged cracks, although other more complex crack configurations are in principle also possible. Now let us introduce a level set function $\phi_i({\bm{\mu}})$ representing the signed distance of a point to the crack in the physical space. In the following two orthogonal level sets are considered to describe the crack and the crack front, such that they satisfy:
\begin{align}
    \phi_1
    (\bm{\mu})  &= ({\bf{F}}(\boldsymbol{\xi}) - {\bf{F}}(\boldsymbol{\xi}_{cr};\bm{\mu})\cdot \mathbf{n}_{cr} ,\\
    \phi_2(\bm{\mu}) &\quad \text{such that} \quad \nabla{\phi_1
    (\bm{\mu}_{ct})}\cdot\nabla{\phi_2 (\bm{\mu}_{ct}) =0, 
    \phi_1(\bm{\mu})} =0, \phi_2(\bm{\mu}_{ct}) =0,
\end{align}
where for the normal level set ${\bf{F}}(\boldsymbol{\xi}_{cr};\bm{\mu})$ denotes a point where the minimum distance to the crack surface is obtained and $\mathbf{n}_{cr}$ is the outward normal vector to the crack surface at that point. The tangential level set refers to the crack tip that is described by the geometric parameters $\bm{\mu}_{ct}$. It is noted that the geometry of the crack in physical space depends on geometric parameters $\bm{\mu}$, thus the LSFs become parameter-dependent. Thereafter, each LSF $\phi_i({\bm{\mu}})$ is discretized using B-spline basis functions as
\begin{equation}\label{levelset_interp}
	\phi_i^h(\boldsymbol{\xi};\bm{\mu})  = \sum_{\bf{j}} \mathcal{B}_{{\bf{j}},p}(\boldsymbol{\xi})\phi_{{i}}^{\bf{j}}({\bm{\mu}}),
\end{equation}
where the vector $\phi_{{i}}^{\bf{j}}({\bm{\mu}})$ contains the coefficients associated to the LSF $\phi_i({\bm{\mu}})$. 

\subsection{Parameterized solution discretization}\label{sec3_2}
This section briefly illustrates the solution discretization in a parameterized setting. 
 To illustrate the imposition of disconituity with enrichment functions, let us first introduce the Heaviside function $H(\bm{\mu})$ as: 
\begin{equation}\label{heaviside}
	H (\bm{\mu})= 
	\begin{cases}
		&+1 \quad \text{if} \quad \phi_1({\bm{\mu}})  > 0\\ 
		&-1  \quad \text{if} \quad  \phi_1({\bm{\mu}})  < 0, 
	\end{cases}
\end{equation}
where $\phi_1({\bm{\mu}})$ refers to the normal LSF representing the signed distance to the crack surface.
Moreover, for the singularity at the crack tip we consider the asymptotic displacement field $\boldsymbol{u}^{\infty}({\bm{\mu}})$ near the crack tip as a function of the polar coordinates $(r({\bm{\mu}}),\theta({\bm{\mu}}))$ according to \cite{Ventura2022,Fleming1997} as
\begin{equation}\label{asymptotic}
	\boldsymbol{u}^{\infty}(\bm{\mu})=\begin{bmatrix}
		\sqrt{r(\bm{\mu})}\text{sin}\frac{\theta(\bm{\mu})}{2}, \sqrt{r(\bm{\mu})}\text{cos}\frac{\theta(\bm{\mu})}{2},
		\sqrt{r(\bm{\mu})}\text{sin}\frac{\theta(\bm{\mu})}{2}\text{sin}\theta(\bm{\mu}), 	\sqrt{r(\bm{\mu})}\text{cos}\frac{\theta(\bm{\mu})}{2}\text{sin}\theta(\bm{\mu})
	\end{bmatrix},
\end{equation}
where $r(\bm{\mu})$ is the distance to the tip, $\theta(\bm{\mu})$ is the angle with respect to the direction of propagation of the crack with the coordinate system being centered at the crack tip. Next, let us demonstrate how to obtain the polar coordinates in the physical space. These can be defined in terms of the LSFs as:
\begin{equation}\label{polar3d}
	r(\bm{\mu}) = \sqrt{\phi_1^h(\boldsymbol{\xi};{\bm{\mu}})^2 + {\phi_2^h(\boldsymbol{\xi};{\bm{\mu}})^2}}, \qquad \theta(\bm{\mu})=\arctan\left(\frac{\phi_1^h(\boldsymbol{\xi};{\bm{\mu}})}{\phi_2^h(\boldsymbol{\xi};{\bm{\mu}})}\right),
\end{equation}
where $\phi_1^h$ and $\phi_2^h$  can be evaluated with Eq.\eqref{levelset_interp}.

Having these definitions at hand, the solution discretization ${\boldsymbol{u}}_h(\boldsymbol{\mu}) \in V_h(\boldsymbol{\mu})$ for the parameterized problem reads: 
\begin{equation}\label{solution}
	{\boldsymbol{u}}_h(\boldsymbol{\mu}) = \sum_{{\bf{i}} \in \mathcal{L}} \mathcal{B}_{\bf{i},p}(\boldsymbol{\xi}) {\bf{U_i}} + H(\bm{\mu}) \sum_{{\bf{i}}' \in \mathcal{H}(\boldsymbol{\mu})} \mathcal{B}_{\bf{i}',p}(\boldsymbol{\xi}) \tilde{\bf{U}}_{\bf{i}'} +  \sum_{{\bf{i}}'' \in \mathcal{T}(\boldsymbol{\mu})} \mathcal{B}_{\bf{i}'',p}(\boldsymbol{\xi}) ( \sum_{j=1}^{4} \boldsymbol{u}^{\infty}_j(\bm{\mu})\bar{{\bf{U}}}_{\bf{i}''}^j),	
\end{equation}
where the discrete subspace $V_h(\boldsymbol{\mu})$ is spanned by the enriched B-spline basis, $\mathcal{L}$ contains the indices of the basis functions corresponding to the continuous part of the solution, $\mathcal{H}(\boldsymbol{\mu})$ the indices of the basis functions that are cut or partially cut by the discontinuity, $\mathcal{T}(\boldsymbol{\mu})$ the indices of the basis functions used for the support of the tip enrichment and ${{\bf{U}}}, \tilde{{\bf{U}}}, \bar{{\bf{U}}} $ contain the respective control variables for the continuous, discontinuity enriched and asymptotically enriched parts of the solution. It becomes clear that the sets of enriched basis functions $\mathcal{H}(\boldsymbol{\mu})$,$\mathcal{T}(\boldsymbol{\mu})$ depend on the choice of the parameters $\bm{\mu}$ describing the crack, and therefore the solution discretization space $V_h(\boldsymbol{\mu})$ is also $\bm{\mu}$-dependent.

With this solution discretization, the discrete approximation of the problem in Eq.~\eqref{weak} leads to the following parameterized linear system of dimension $\mathcal{N}_h(\bm{\mu}) = \text{dim}(V_h(\bm{\mu}))$
\begin{equation}\label{eqsystem}
	{\bf{A}}(\bm{\mu}){\bf{u}}_h(\bm{\mu}) = {\bf{f}}(\bm{\mu}),
\end{equation}
where ${\bf{A}}({\bm{\mu}})$ $\in \mathbb{R}^{\mathcal{N}_h(\bm{\mu})\times \mathcal{N}_h(\bm{\mu})}$ is the stiffness matrix, ${\bf{f}}({\bm{\mu}}) \in \mathbb{R}^{\mathcal{N}_h(\bm{\mu})}$ is the vector representing the body force and $\mathcal{N}_h(\bm{\mu})$ is the number of degrees of freedom. The problem in Eq.\eqref{eqsystem} is the high-fidelity or full order model (FOM) forming the basis to construct a parametric ROM. 

\section{Local non-intrusive reduced basis method}\label{sec4}
This section presents a localized strategy to construct efficient ROMs for the problem at hand. The starting point is a many-query context, where multiple solution evaluations are required for the high-fidelity problem in Eq.~\eqref{eqsystem} and for different values of the parameters $\boldsymbol{\mu}$. This is, for example, the case with crack detection tasks: given some actual data such as measurements from installed sensors, crack detection entails the solution of an inverse problem. This requires multiple solves of the high-fidelity model to identify the location of a crack. The use of ROMs can therefore reduce the computational effort and speedup the solution of such parameterized problems. The latter can be tackled in a fast manner by establishing an efficient offline/online split of the simulation process. In the offline phase, several solution snapshots of the high-fidelity model are computed to construct a reduced basis based on techniques such as the {POD}. This step results in a linear combination of reduced basis functions that is computed offline once and for all. Then, a non-intrusive ROM can be constructed by orthogonal projection of the high-fidelity snapshots onto the reduced basis subspace and nonlinear regression via neural networks. This allows the treatment of non-affine problems in an non-intrusive fashion. The main advantage of such ROMs is that they fully exploit the solution snapshots generated for the construction of the reduced basis in the offline phase and require minimal or no modifications to the underlying high-fidelity solvers. In the online phase, the evaluation of the constructed ROM allows to obtain the solution rapidly for any given value of the parameter. More details on the reduced basis method and its non-intrusive variants are provided in \cite{Hesthaven2016,QMN_RBspringer} and \cite{Hesthaven2018,Chen2021,DalSanto2020}, respectively. In particular, the presence of geometric parameters and discontinuities such as cracks poses the following challenges to constructing efficient ROMs for parameterized problems: 
\begin{itemize}
\item The enriched space $V_h({\boldsymbol{\mu}})$ and its dimension $\mathcal{N}_h(\boldsymbol{\mu})$ depend on geometric parameters that describe the crack. Namely, the sets of basis functions $\mathcal{H}({\boldsymbol{\mu}})$, $\mathcal{T}({\boldsymbol{\mu}})$ that are cut or partially cut by the discontinuity are $\boldsymbol{\mu}$-dependent and change from one snapshot to the other. This requires suitable processing and mapping of solutions prior to forming a snapshots matrix and constructing a reduced basis.
\item The problems at hand involve geometric parameters and are in general non-affine. Moreover, the solution manifold is obtained from geometric variations representing moving discontinuities and may be therefore highly nonlinear with respect to the parameters $\boldsymbol{\mu}$. The application of a a standard, global ROM to approximate the nonlinear manifold may result in a very large dimension of the basis and poor approximation properties.
\end{itemize}
In what follows, we will elaborate on a localization strategy to construct efficient ROMs for such parameterized problems. Instead of relying on a single linear reduced basis space, the idea is to construct multiple local subspaces to approximate different regions of the solution space in a piecewise linear fashion. This can be achieved using clustering and classification strategies, while relying on local non-intrusive ROMs to approximate the underlying problem in an efficient manner. The interested reader is further referred to previous works on localized ROMs for, e.g., nonlinear systems \cite{Amsallem2012,Geelen2022} and isogeometric analysis \cite{Chasapi2023,Chasapi2024}. The main steps involved in the ROM approach that will be elaborated in the next sections are summarized as follows:
\begin{itemize}
\item Snapshots computation and mapping: A snapshots matrix is computed offline for a given set of parameters by solving the FOM. Each snapshot is mapped onto a common mesh configuration prior to constructing a reduced basis.
\item Clustering strategy: A clustering strategy is applied to partition the snapshots matrix based on the similarities between solution snapshots. Each parameter is then assigned to the same cluster as its corresponding snapshot. 
\item Classification procedure: We train a classifier offline that maps each parameter to its corresponding cluster.
\item Proper Orthogonal Decomposition: Local reduced bases are constructed for each cluster by applying the POD algorithm.
\item Local reduced basis approximation: We construct local non-intrusive ROMs for each cluster. The solution snapshots are projected onto the reduced basis space for each cluster. Then neural networks are trained offline to learn the map between a parameter and its corresponding reduced coefficients.
\item Online phase: Given a new parameter in the online phase, we evaluate the classifier to select the most suitable cluster and the neural network to obtain an approximation of the reduced coefficients. The approximation of the solution is obtained by projection using the reduced basis constructed in the offline phase.
\end{itemize}

\subsection{Snapshots computation and mapping}\label{sec:snapshots}
\begin{figure}[!h]
	\centering
	\includegraphics[width=0.6\textwidth]{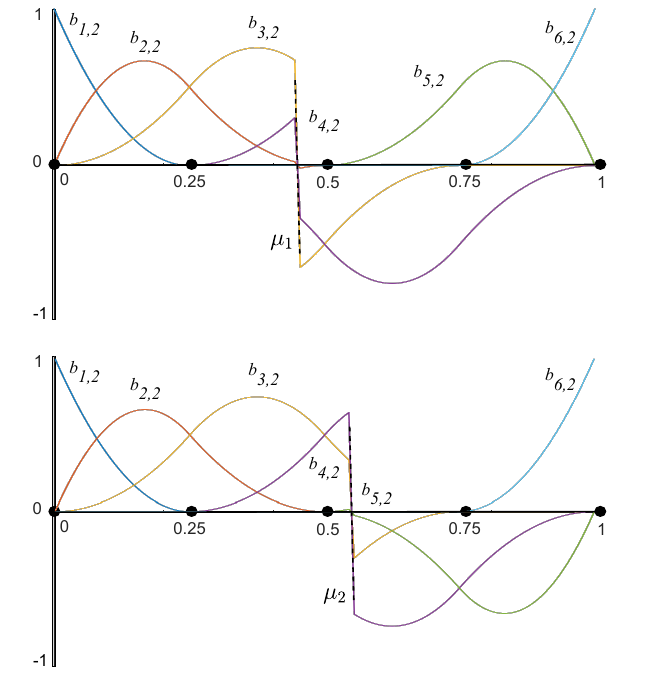} \\
	\caption{Univariate B-spline basis for two different parameters $\mu_1$ and $\mu_2$ describing the location of the crack.}\label{fig:Bspline}
\end{figure}

Let us first consider the solution of parameterized problems on spline discretizations using XIGA. Since the crack is incompatible with the mesh and modelled through discontinuous enrichment functions, a strong discontinuity is introduced in the solution field. As the crack depends on the parameters $\boldsymbol{\mu}$, the spline enrichment functions become also parameter dependent. Fig.~\ref{fig:Bspline} illustrates a representative example for a univariate B-spline basis. Given a parameter $\mu$ that describes the location of the crack, the basis functions affected by the discontinuity are highlighted by the vertical dashed line in Fig.~\ref{fig:Bspline}. Clearly, the basis functions that are cut or partially cut by the discontinuity change from one parameter to the other. 
In fact, the enriched spline space $V_h({\boldsymbol{\mu}})$ and its dimension $\mathcal{N}_h(\boldsymbol{\mu})$ are $\boldsymbol{\mu}$-dependent. This requires further measures prior to constructing a snapshots matrix, as the number of enriched basis functions and thus the length of each solution vector may differ from one crack configuration to the other. Therefore, a suitable mapping of the solution vector $\mathbf{u}_h(\boldsymbol{\mu})$ is necessary. Several choices are in principle possible \cite{Karatzas2020}.
In fact, the following mapped snapshot can be defined by composition for $\boldsymbol{\mu} \in \mathcal{P}$ as
\begin{equation}\label{transport}
	\tilde{\mathbf{u}}(\boldsymbol{\mu}) = \mathbf{u}_h(\boldsymbol{\mu}) \circ \ \boldsymbol{\tau} (\boldsymbol{\mu}),
\end{equation}
where $\boldsymbol{\tau} (\boldsymbol{\mu})$ is assumed to be a bijective, invertible map and  $\tilde{\mathbf{u}}(\boldsymbol{\mu})$ the mapped solution that can be used for the construction of a snapshots matrix. In this work, a mesh representation composed by the image of the knots in the physical space forms the basis for the above mapping. For the continuous part of the solution in Eq.~\eqref{solution}, this involves evaluating a spline mapping in line with Eq.~\eqref{splinemap}. Since the crack is defined by level sets in the physical space, the above map can be easily obtained with simple algebraic manipulations for the enriched part of the solution. Note that this approach bears connections with snapshots preprocessing based on transportation \cite{Cagniart2019,Nair2019}, for example, via problem specific transformations \cite{Karatzas2020}, shifted POD modes \cite{Reiss2018} or optimal transport \cite{Bernard2018}.

\begin{remark}
It should be noted that the mapping involves only a postprocessing of the solutions that are obtained from different crack configurations without modifications in the high-fidelity solvers. Therefore, it allows a seamless integration with non-intrusive ROM techniques. 
\end{remark}

\begin{remark}
It is assumed that the domain itself is parameter independent and the geometric parameters of interest are only related to the definition of the crack. By introducing the snapshots mapping, we construct a reduced basis that is defined on a $\boldsymbol{\mu}$-independent computational domain. In a similar fashion, the solution of the problem is also expressed on a $\boldsymbol{\mu}$-independent domain. Its values for a given crack configuration can be obtained by evaluating the inverse of the above mapping. 
\end{remark}

\subsection{Clustering strategy}\label{sec:cluster}
In the following, we will elaborate on the clustering strategy to construct a local ROM. The idea is to construct multiple local reduced bases instead of one global basis of high dimension by partitioning snapshot solutions in the offline phase. In principle, different choices are possible for the clustering algorithm \cite{Amsallem2012,Haasdonk2011,Chasapi2023}. Typically, hard clustering algorithms partition snapshots into distinct clusters \cite{Likas2003,Xu2005}. In what follows, we opt for snapshot-based clustering with the fuzzy c-means (FCM) clustering algorithm \cite{Gosh2013, Geelen2022}. Based on the FCM algorithm, each snapshot can be assigned to multiple clusters according to a varying degree of membership. The main idea is to assign a given solution $\tilde{\mathbf{u}}({\boldsymbol{\mu}})$ to the cluster that minimizes the weighted distance $\mathcal{D}({\bm{\mu}},k)$ between the snapshot itself  and a cluster center, given for $\boldsymbol{\mu} \in \mathcal{P}$ as

\begin{equation}\label{cdistance}
	\mathcal{D}({\bm{\mu}},k) = \omega_k  \norm{{\tilde{\mathbf{u}}({\boldsymbol{\mu}}) - {{\boldsymbol{c}}}_k}}_2^2, \quad k=1,\dots,N_c.
\end{equation}
Here, $N_c$ is the number of clusters, ${\boldsymbol{c}_k}$ denotes the cluster center of the $k$-th cluster and $\omega_k$ is the weight representing the degree to which the given snapshot $\tilde{\mathbf{u}}({\boldsymbol{\mu}})$ belongs to the $k$-th cluster.\footnote{\label{note1}For the sake of conciseness, the range $k=1,\dots,N_c$ of the cluster index may be omitted in the following.}

In order to perform the clustering, we create a sufficiently fine and properly selected training sample set $\mathcal{P}_{train}=\{\bm{\mu}_1,\dots,\bm{\mu}_{N_s}\} \subset \mathcal{P}$ of dimension $N_s=\text{dim}(\mathcal{P}_{train})$. Thereafter, one can form the solution snapshots matrix $\mathbb{S} \in \mathbb{R}^{\tilde{\mathcal{N}}_h \times N_s}$
\begin{equation}\label{Smatrix}
\mathbb{S} = [\tilde{\mathbf{u}}_1, \tilde{\mathbf{u}}_2, \dots, \tilde{\mathbf{u}}_{N_s}],
\end{equation}
where the vectors $\tilde{\mathbf{u}}_j \in \mathbb{R}^{{\tilde{\mathcal{N}}}_h}$ represent the mapped solutions $\tilde{\mathbf{u}}(\boldsymbol{\mu}_j)$ of dimension $\tilde{\mathcal{N}}_h$ for $j=1,\cdots,N_s$. In the following, we seek a partition of the snapshots matrix $\mathbf{S}=\lbrace\mathbb{S}_1, \mathbb{S}_2,\dots, \mathbb{S}_{N_c}\rbrace$ using the FCM algorithm.
The cluster centers, i.e. centroids, are chosen randomly at the beginning and updated iteratively until the algorithm converges such that
\begin{equation}\label{cluster_center}
{{\boldsymbol{c}}}_k = \frac{\sum_{j=1}^{N_s}\omega_k^j  {\tilde{\mathbf{u}}_j}} {\sum_{j=1}^{N_s}\omega_k^j}, \quad k=1,\dots,N_c.
\end{equation}
Note that the algorithm requires to chose in advance the number of clusters $N_c$, and a suitable choice is in principle problem dependent. Moreover, the membership degree associated to the $j$-th snapshot is updated iteratively such that
\begin{equation}\label{cluster_weight}
	\omega_k^j = \left(\sum_{l=1}^{N_c}\frac{\norm{{\tilde{\mathbf{u}}_j}-{{\boldsymbol{c}}}_k}}{\norm{{\tilde{\mathbf{u}}_j}-{{\boldsymbol{c}}}_l}}\right)^{-\frac{2}{r-1}} \quad j=1,\dots,N_s.
\end{equation}
Here, $r$ denotes the fuzzy partition matrix exponent that controls the overlap at the boundaries between clusters being $r > 1$. In the numerical experiments of this work the exponent is set to $r=1.5$. It should be noted that in contrast to localized ROMs based on $\textit{k-means}$ clustering, one may obtain overlapping clusters with smooth transitions from one cluster to another without further algorithmic treatment \cite{Amsallem2012}. In principle, a snapshot can be assigned to a given cluster based on the degree of membership. In this work, the partitioning is performed such that $\forall \bm{\mu} \in \mathcal{P}_{train}$
\begin{equation}\label{partition}
	{\mathbb{S}_k}= \{\tilde{\mathbf{u}}({\bm{\mu}}) \ |\:\ \text{if}\  \underset{i=1,\dots,N_c}{\arg \max} \omega_i=k\}.
\end{equation}
 
\begin{remark}
In the numerical experiments of this work the clustering is performed using the high-fidelity solution vectors. Nevertheless, it can be advantageous to reduce the dimension of the snapshots prior to applying the clustering algorithm \cite{Geelen2022, Kaiser2014}. This is particularly relevant for high-dimensional problems, such as in a time-dependent,  three-dimensional setting, and may increase the stability of clustering. 
\end{remark}
Once the snapshots are partitioned, we consider a partitioning of the parameter space $\mathcal{P} = \bigcup_{k=1}^{N_c}\mathcal{P}^k$ into $N_c$ subspaces. Note that this forms the basis for the classification procedure, which will be described in Sec.~\ref{sec:classify}. 
To perform the partitioning, the training set $\mathcal{P}_{train}$ can be expressed as
\begin{equation}\label{partition2}
\mathcal{P}_{train} = \bigcup_{k=1}^{N_c}\mathcal{P}^k_{train}
\end{equation}
where $\mathcal{P}^k_{train}$ contains the parameters associated to the snapshots of the submatrix ${\mathbb{S}_k}$. This is motivated by the fact that each snapshot is associated to one parameter vector $\bm{\mu} \in \mathcal{P}_{train}$, therefore a given parameter can be attributed to the same cluster as its corresponding solution snapshot.

\subsection{Classification procedure}\label{sec:classify}
The idea behind local ROMs is to train multiple local reduced basis approximations in the offline phase and then select a suitable ROM on the fly for a given value of the parameters. In this work, we will opt for classification to facilitate the selection process. The reader is also referred to classification-based ROM selection for nonlinear PDEs in the framework of localized DEIM approximations \cite{Peherstorfer2014} and non-intrusive ROMs \cite{Geelen2022}. In principle, different indicators are possible to train the classifier in the offline phase. In what follows, we will employ parameter-based classification for the selection of the most suitable ROM. 

Let us assume that a partitioning of the snapshot matrix in Eq.~\eqref{Smatrix} is given based on the clustering strategy in Sec.~\ref{sec:cluster}. The parameter-based classification is motivated by the fact that each snapshot $\tilde{\mathbf{u}}(\bm{\mu})$ is associated to one parameter vector $\bm{\mu} \in \mathcal{P}$. Therefore, each parameter vector can be assigned to the same cluster as its corresponding snapshot and the cluster index itself can serve as label for training a classifier. In what follows, we seek for an approximation of the classifier $\mathcal{C}$ using the parameter vector $\bm{\mu} \in \mathcal{P}$ as indicator such that 
\begin{equation}\label{classifier}
	\hat{\mathcal{C}} \approx \mathcal{C} \ :\ \mathcal{P} \to \{1,2,\dots,N_c\}.
\end{equation}
To perform the classification, we consider again the training sample set $\mathcal{P}_{train} \subset \mathcal{P}$ employed for the clustering in Section~\ref{sec:cluster}. In this work, we will use a nearest neighbor classifier with a Minkowski distance metric although other choices are in principle also possible. The distance is defined in the following as \cite{Xu2005}:
\begin{equation}\label{eqMink}
	d_{j} = \big(\sum_{j=1}^{N_s} |{\mu}_{ij}-k_{j}|^{\frac{1}{q}}\big)^q , \ i=1,\dots,P,
\end{equation}
where $k_j$ denotes the cluster index associated to the $j$-th parameter sample and $q$ is the distance exponent. In the numerical experiments of this work the exponent is set to $q=2$. Once the clustering has been performed, the classifier can be trained in the offline phase. Given any parameter $\bm{\mu} \in \mathcal{P}$ in the online phase, the cluster index $\bar{k} \in \{1,\dots,N_c\}$ is simply obtained by evaluating the constructed classifier $\hat{\mathcal{C}}$ as
\begin{equation}\label{eqC}
	\hat{\mathcal{C}}(\bm{\mu}) = \bar{k}.
\end{equation}

\subsection{Proper Orthogonal Decomposition}\label{sec:pod}
In the following, we briefly review the construction of the reduced basis using the POD. The latter is based on the singular value decomposition algorithm (SVD) and aims to reduce the dimension by extracting a set of orthonormal basis functions \cite{Quarteroni2017}. 
The SVD of a matrix ${\mathbb{S}} \in \mathbb{R}^{m \times n}$ can be expressed as:
\begin{equation}\label{svd}
	{\mathbb{S}}= \mathbb{U}\boldsymbol{\Sigma}\mathbb{Z}^T,
\end{equation}
where the matrices $\mathbb{U}=[\bm{\zeta}_1,...,\bm{\zeta}_{m}] \in \mathbb{R}^{m \times m}$, $\mathbb{Z}=[\bm{\psi}_1,...,\bm{\psi}_{n}] \in \mathbb{R}^{n \times n}$ are orthogonal with columns containing the left and right singular vectors of ${\mathbb{S}}$, accordingly, while the rectangular diagonal matrix $\boldsymbol{\Sigma}=\text{diag}(\sigma_1,...,\sigma_r)\in \mathbb{R}^{{m \times n}}$ contains the singular values $\sigma_1 \ge \sigma_2 \ge ... \sigma_r$, and $r \le \text{min}(m,n)$ is the rank of $\mathbb{S}$. We can thereafter define the POD basis of dimension $N$ as the set of first $N$ left singular vectors of $\mathbb{S}$ corresponding to the $N$ largest singular values as
\begin{equation}\label{basis}
	\mathbb{V} = [\bm{\zeta}_1,...,\bm{\zeta}_{N}] \in \mathbb{R}^{m \times N}.
\end{equation}
The POD basis is orthonormal by construction and its dimension $N$ can be selected such that the error in the POD basis is smaller than a prescribed tolerance $\varepsilon_{POD}$ \cite{QMN_RBspringer}. Thus, $N$ is considered as the smallest integer such that
\begin{equation}\label{tolerance}
	1 - \frac{\sum_{i=1}^{N}\sigma_i^2}{\sum_{i=1}^{r}\sigma_i^2} \le \varepsilon_{POD}^2.
\end{equation}
Since the error entails the sum of the squares of the singular values associated to the neglected POD modes, this ensures that the energy captured by the last neglected modes is smaller than or equal to $\varepsilon_{POD}$. 

\subsection{Local reduced basis approximation}\label{sec:localRB}
Let us now illustrate the localized approach to construct efficient ROMs in a non-intrusive manner. In this work, we use artificial neural networks to construct non-intrusive ROMs \cite{Hesthaven2018,Chen2021} although other options are in principle also possible as elaborated in \cite{Georgaka2020,Benner2021,Ghattas2021} and the references therein.
Based on the clustering strategy discussed in Section \ref{sec:cluster}, a separate reduced basis can be constructed offline for each cluster. In what follows we seek for local reduced bases ${\mathbb{V}}_k \in \mathbb{R}^{\tilde{\mathcal{N}}_{h}\times N_k}$ of dimension $N_k$ for each cluster $k=1,\dots,N_c$ and use the POD for the construction of the bases as discussed in Section~\ref{sec:pod}. Note that the dimension of the problem at hand can be sufficiently reduced by constructing multiple local subspaces instead of one global reduced basis, that is $N_k \ll \tilde{\mathcal{N}}_{h}$. 

In order to construct a reduced basis using the POD, let us again consider a fine training sample set $\mathcal{P}_{train} = \{\bm{\mu}_1,...,\bm{\mu}_{N_s}\} \subset \mathcal{P}$ of dimension $N_s = \text{dim}(\mathcal{P}_{train})$ as introduced in Section~\ref{sec:cluster} and the snapshots matrix $\mathbb{S}$ in Eq.~\eqref{Smatrix}.
After partitioning the snapshots into $N_c$ submatrices as $\mathbf{S}=\lbrace {\mathbb{S}}_1,\dots,{\mathbb{S}}_{N_c}\rbrace$, the local POD basis ${\mathbb{V}}_k$ can be extracted seperately from each submatrix for $k=1,\dots,N_c$ according to Section~\ref{sec:pod}. Thus, the local POD basis reads:
\begin{equation}\label{eq60}
	{\mathbb{V}}_k = [\bm{\zeta}_1^k,\dots,\bm{\zeta}_{N_k}^k] \in \mathbb{R}^{\tilde{\mathcal{N}}_{h} \times N_k}.
\end{equation}
Having these definitions at hand, we can now express the local reduced basis approximation. We seek for an approximation of the solution $\tilde{{\bf{u}}}(\bm{\mu})$ using the local reduced basis ${\mathbb{V}}_k$ for any ${\bm\mu} \in \mathcal{P}$ such that
\begin{equation}\label{eq52}
	\tilde{{\bf{u}}}(\bm{\mu}) \approx  {\mathbb{V}}_k {\bf{u}}_{N}(\bm{\mu}),
\end{equation}
where ${\bf{u}}_N(\bm{\mu})\in \mathbb{R}^{N_k}$ is the reduced solution vector for our problem and the index $k$ is obtained by evaluating Eq.~\eqref{eqC}. Hence, it holds
\begin{equation}\label{eq53}
		\tilde{{\bf{u}}}_h(\bm{\mu}) = \sum_{i=1}^{N_k} ({\mathbb{V}}_k^T 	\tilde{{\bf{u}}}(\bm{\mu}))_i \bm{\zeta}_i^k,
\end{equation}
where $\tilde{{\bf{u}}}_h(\bm{\mu})$ can be viewed as an approximation of the high-fidelity problem.
Once a local reduced basis is constructed, we aim to approximate the map from a parameter vector $\bm{\mu} \in \mathcal{P}^{k}$ to the coefficients ${\mathbb{V}}_k^T \tilde{{\bf{u}}}(\bm{\mu})$ in the expansion of $\tilde{{\bf{u}}}_h(\bm{\mu})$ as
\begin{equation}\label{eq54}
\hat{\mathcal{Z}}_k \approx \mathcal{Z}_k \ :\ \bm{\mu} \in \mathcal{P}^{k} \subset \mathbb{R}^P  \to {\mathbb{V}}_k^T \tilde{{\bf{u}}}(\bm{\mu}) \in \mathbb{R}^{N_k}.
\end{equation} 
Note that the map is assumed to be sufficiently smooth. Given a new parameter $\bm{\mu} \in \mathcal{P}$, we can obtain an approximation of the reduced solution by evaluating the map at the parameter $\bm{\mu}$ as
 \begin{equation}\label{eq55}
{ {\bf{u}}}_N(\bm{\mu}) \approx	{\hat{\bf{u}}}_N(\bm{\mu})=\hat{\mathcal{Z}}_k(\bm{\mu}),
 \end{equation} 
where the index $k$ is obtained by evaluating the classifier in Eq.~\eqref{eqC}. Once the reduced solution is obtained, the approximation of the solution can be constructed by evaluating Eq.~\eqref{eq52}. Given that the size of the reduced approximation is small, i.e., $N_k \ll \tilde{\mathcal{N}}_{h}$, it is suitable for multiple online evaluations for different values of the parameters $\bm{\mu}$. It should be noted that the accuracy of the reduced basis solution depends highly on the quality of the reduced basis and the approximation in Eq.~\eqref{eq54}.

\begin{figure}[!h]
	\centering
	\includegraphics[width=0.6\textwidth]{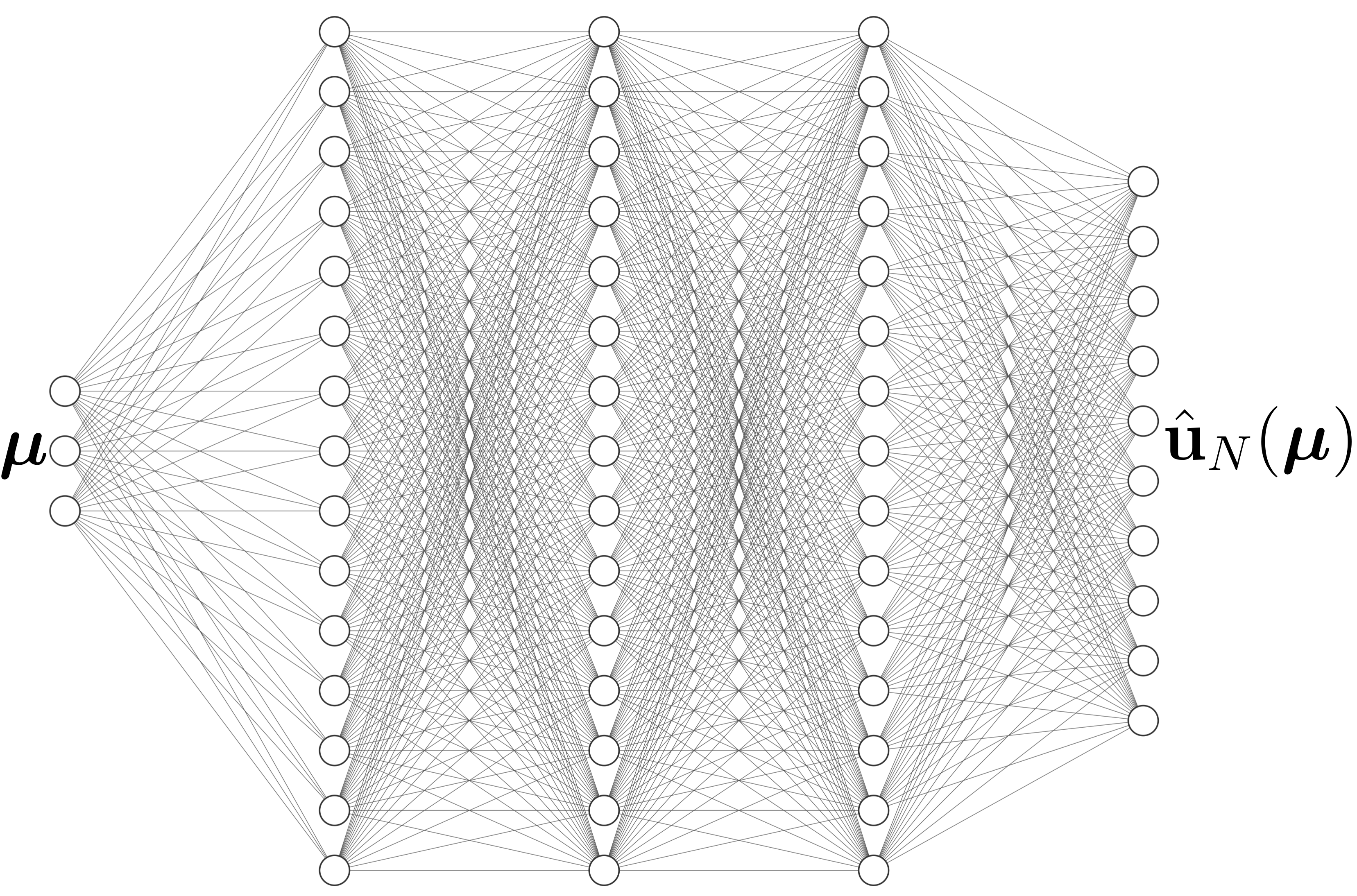} \\
	\caption{Approximation of the map in Eq.~\eqref{eq54} via neural networks.}\label{fig:network}
\end{figure}

To construct such an approximation, we consider a fine training sample set $\mathcal{P}_{train} = \{\bm{\mu}_1,...,\bm{\mu}_{N_s}\} \subset \mathcal{P}$ of dimension $N_s = \text{dim}(\mathcal{P}_{train})$ as introduced previously. In the offline phase, the training set is first clustered based on the strategy in Section \ref{sec:cluster}. Next, we construct local reduced bases separately for each cluster and train neural networks to approximate the map in Eq.~\eqref{eq54}. Given a new parameter $\bm{\mu} \in \mathcal{P}$ in the online phase, the network is evaluated to obtain an approximation of the reduced solution in Eq.~\eqref{eq55} as depicted in Fig.~\ref{fig:network}. In the numerical experiments of this work, the same snapshots generated for the construction of the reduced basis are also used for learning the map in Eq.~\eqref{eq54}, although this choice is in principle flexible.


The training is performed such that for any $\bm{\mu} \in \mathcal{P}_{train}^k$ it minimizes a cost function   
\begin{equation}\label{eq57}
	\min_{\boldsymbol{\theta}}\mathcal{L}_k \left( \;\hat{\mathcal{Z}_k}(\boldsymbol{\theta};\bm{\mu}),\;\mathbb{V}_k^T \tilde{{\bf{u}}}(\bm{\mu}) \;\right),
\end{equation}
where $\boldsymbol{\theta}$ denotes the parameters of the model, i.e., the weights and biases. In the numerical experiments of this work, multi-layer perceptrons with $L$ hidden layers and $H$ neurons per layer are employed for the training. The cost function is chosen such that for $\bm{\mu} \in \mathcal{P}_{train}^k$ the mean squared error is minimized during the training process with a regularization potential as 
\begin{equation}\label{eq57}
\mathcal{L}_k =	 \mathcal{L}_{data}(\boldsymbol{\theta}) + \mathcal{R}(\boldsymbol{\theta})= \norm{{{\bf{u}}}_{NN}(\boldsymbol{\theta};\bm{\mu})-\mathbb{V}_k^T \tilde{{\bf{u}}}(\bm{\mu})}^2 + \frac{\lambda}{2}\norm{\boldsymbol{\theta}}^2,
\end{equation}
where $\mathcal{R}(\boldsymbol{\theta})$ is the regularization potential, $\lambda$ is the regularization parameter  and ${{\bf{u}}}_{NN}(\boldsymbol{\theta};\bm{\mu}) \in \mathbb{R}^{N_k}$ is the actual output provided by the network. Before performing the training, feature scaling is applied to adjust the range of input and output \cite{Chen2021}. For the input, the minimum and maximum values of the parameters are used for the scaling such that for any ${\bm{\mu}} \in \mathcal{P}_{train}^k$
\begin{equation}\label{eq58}
	{\bm{\mu}}^{s}=\frac{\bm{\mu}-\bm{\mu}_\text{min}}{\bm{\mu}_\text{max}-\bm{\mu}_\text{min}}.
\end{equation}
For the output, we rely on normalization based on the mean and standard deviation computed from all output data. Given a parameter $\bm{\mu} \in \mathcal{P}_{train}^k$, the output data used for training are rescaled as
\begin{equation}\label{eq59}
	{\bf{u}}_N^{s}(\bm{\mu})=\frac{{\tilde{\bf{u}}}_N(\bm{\mu})-\bar{\bf{u}}_N}{\sigma_{\tilde{\bf{u}}_N}},
\end{equation}
where ${\tilde{\bf{u}}}_N(\bm{\mu})=\mathbb{V}_k^T \tilde{{\bf{u}}}(\bm{\mu})$ are the reduced coefficients,
$\bar{\bf{u}}_N$ is the mean and ${\sigma_{\tilde{\bf{u}}_N}}$ is the standard deviation computed from the data set. Once the network is trained, the final prediction can be obtained for any $\bm{\mu} \in \mathcal{P}$ as
\begin{equation}\label{eq60}
	\hat{\textbf{u}}_{N}(\bm{\mu})={\bf{u}}_{NN}^{s}({\bm{\mu}}) {\sigma_{\tilde{\bf{u}}_N}}+\bar{\bf{u}}_N,
\end{equation}
where ${{\bf{u}}}^s_{NN}(\bm{\mu}) \in \mathbb{R}^{N_k}$ is the rescaled output provided by the network. The offline procedure to construct the localized ROM and the online phase are presented in Algorithms 1 and 2, respectively.
\makeatletter
\def\algbackskip{\hskip-\ALG@thistlm}
\makeatother
 \begin{algorithm}
    \caption{Offline phase}\label{algo1}
    \begin{algorithmic}[1]
    \Procedure{[$\hat{\mathcal{C}},\{\hat{\mathcal{Z}_k}\}_{k=1}^{N_{c}}$] = OFFLINE}{${\mathcal{P}_{train}}, N_c, \epsilon_{POD}$}
     \For{${\bm{\mu}} \in \mathcal{P}_{train}$} 
    \State \emph{Full order model}:
    \State ${\bf{u}}_h{(\bm{\mu})} \gets \textit{solve FOM in } \eqref{eqsystem}$ 
    \State ${\tilde{\bf{u}}}{(\bm{\mu})} \gets \textit{map solution vector in } \eqref{transport}$ 
    \State \emph{Solution snapshots}:
    \State ${\mathbb{S}} = [{\mathbb{S}}, \tilde{\bf{u}}{(\bm{\mu})}]$
    \EndFor
    \State $\{\mathbb{{{S}}}_k\}_{k=1}^{N_{c}} \gets \textit{fuzzy c-means clustering } (\mathbb{S},N_{c})$
     \State $\{{\mathcal{P}^1_{train},.., \mathcal{P}^{N_{c}}_{train}}\} \gets \textit{partition parameters} \ \mathcal{P}_{train} \ \eqref{partition2}$ 
    \State $\hat{\mathcal{C}} \gets \textit{nearest neighbor classifier } (\{{\mathcal{P}^1_{train},.., \mathcal{P}^{N_{c}}_{train}}\},\{1,\dots,N_{c}\})$
     \State \emph{Local reduced basis functions and approximations}:
    \For{$k = 1,...,N_c$}    
     \State ${\mathbb{V}}_k \gets \textit{POD}({\mathbb{S}}_{k},\epsilon_{POD});$
     \State $\mathcal{P}^{k,s}_{train},{\mathbb{S}}_{k}^s \gets \textit{feature scaling} \ \eqref{eq58}-\eqref{eq59}$ 
      \State $\hat{\mathcal{Z}_k} \gets \textit{construct approximation } (\mathcal{P}^{k,s}_{train},{\mathbb{V}}_k^T{\mathbb{S}}_{k}^s) \ \eqref{eq54}$
    \EndFor
    \EndProcedure
    \end{algorithmic}
    \end{algorithm}

  

\makeatletter
\def\algbackskip{\hskip-\ALG@thistlm}
\makeatother
 \begin{algorithm}
    \caption{Online phase}\label{algo2}
    \begin{algorithmic}[1]
   \Procedure{[${\hat{\bf{u}}}_N$] = ONLINE}{$\hat{\mathcal{C}},\{\hat{\mathcal{Z}_k}\}_{k=1}^{N_{c}},{\bm{\mu}}$}
    \State $\bar{k} =\hat{\mathcal{C}}({\bm{\mu}}) \ \eqref{eqC}$ 
    \State \emph{Reduced order solution}:
    \State ${{\bf{u}}_{NN}^s} = {\hat{\mathcal{Z}}_{\bar{k}}}({\bm{\mu}})$ 
    \State ${\hat{\bf{u}}}_N \gets \textit{rescale network output} \ \eqref{eq60}$ 
    \EndProcedure
    \end{algorithmic}
    \end{algorithm}   

Alternatively, interpolation can be adopted to learn the map between parameters and reduced coefficients. In the following we briefly introduce radial basis functions (RBFs) for the interpolation \cite{Buhmann2003} that will be used later on in the numerical experiments of this work. RBFs are chosen here as they are able to interpolate scattered data, although using other functions (e.g. splines) is also possible. In the following we seek for an approximation of the reduced solution as
\begin{equation}\label{eq20_5}
\mathbf{u}_N(\bm{\mu}) \approx \sum_{j=1}^{N_{s}^k} \omega_{j}^k \phi_{j}^k (\norm{\bm{\mu}-\bm{\mu}_j^k}),
\end{equation}
where $\phi_{j}^k$ is the radial basis function associated to the $j$-th center parameter point $\bm{\mu}_j^k$ of the $k$-th cluster, and the index $k$ is obtained by evaluating the classifier in Eq.~\eqref{eqC}. Note that $N_s^k=\text{dim}(\mathcal{P}_{train}^k)$, while the number of interpolation points coincides with the number of training parameter samples $N_{s}^k$. In order to evaluate the above expression online, the unknown weights $\omega_{j}^k$ are pre-computed offline such that they fulfill the interpolation constraint exactly for ${\bm{\mu}_l} \in \mathcal{P}_{train}^k$
\begin{equation}\label{eq20_4}
\sum_{j=1}^{N_{s}^k} \omega_{j}^k \phi_{j}^k (\norm{\bm{\mu}_l-\bm{\mu}_j^k}) = \mathbf{u}_N(\bm{\mu}_l), \quad l=1,\dots,N_{s}^k, k=1,\dots,N_c.
\end{equation}
In the numerical experiments of Section \ref{sec5} cubic RBFs are employed, although other choices are also possible. Depending on the choice of RBF, the above problem becomes uniquely solvable by augmenting the above definition with polynomials \cite{Buhmann2000, Powell1987, Powell1992}. It should be noted that with increasing number of interpolation points the condition number of the matrix associated to the RBF problem grows as well. However, the above construction of local approximations is based on partitioned parameter sets such that the number of interpolation points is confined for the cluster at hand.


\section{Numerical results}\label{sec5}
This section presents several numerical experiments for linear elastic problems to illustrate the capabilities of the presented reduction strategy. In what follows we consider: (i) a 2D plate with edge crack and a two-dimensional parameterization, where the length and location of the crack is changing, (ii) a 2D plate with center crack and a three-dimensional parameterization, where the length, location and orientation of the crack is changing and (iii) a 3D plate with edge crack and parameterized location of the crack to demonstrate the applicability of the ROM approach to three-dimensional geometries. In all cases we assess the accuracy and efficiency in terms of computational times and speedup. In particular, the accuracy is analyzed for a new parameter $\bm{\mu} \in \mathcal{P}_{test} \subset \mathcal{P}$, where ${\mathcal{P}_{test}}$ is a sufficiently large and randomly selected test sample. The numerical experiments are carried out using the open-source Matlab isogeometric library migfem \cite{Nguyen2015} for the computation of the high-fidelity problem in combination with the open-source library redbKIT \cite{redbKIT} to construct the ROM. Note that unless stated otherwise, we employ Latin hypercube sampling \cite{McKay1979} to construct the parameters for the training and test sets. Moreover, the Levenberg-Marquardt algorithm \cite{Levenberg1944,Marquardt1963} is used to iteratively adjust the weights of the neural networks with either $L^2$ or Bayesian regularization \cite{MacKay1992} unless stated otherwise. For the Levenberg-Marquardt algorithm, the initial value of the bias parameter is set to $\mu = 0.001$ and an early stopping criterion is adopted to further monitor validation loss, that is the training is stopped when the validation error increases over 6 consecutive epochs. Unless stated otherwise, 10$\%$ of the training samples are employed for validation. In general, multiple initializations are performed to limit the effect of random initial guesses and the dependence of the results on the initial configuration at the beginning of the learning process. For our numerical experiments we employ neural networks with the same number of neurons for the hidden layers and the sigmoid function as activation function for the hidden neurons unless stated otherwise. Our numerical investigations show that architectures with up to 5 hidden layers and 15 neurons per hidden layer produce nearly optimal results. The normalization of the input and output data is performed as described in Sec.~\ref{sec:localRB} unless stated otherwise. For all test cases, the tolerance of the POD is chosen as $\epsilon_{POD}=10^{-5}$. The accuracy of the ROM is analyzed for a new parameter $\bm{\mu} \in \mathcal{P}_{test}$, while the dimension of the test sample is $N_t=\text{dim}({\mathcal{P}_{test}})$. The error is analyzed based on the relative average error $\varepsilon_{rel}$ and maximum error $\varepsilon_{max}$ with respect to the high-fidelity solution as:
\begin{equation}
	\varepsilon_{rel}=\frac{1}{N_t}\sum_{i=1}^{N_t}
	\frac{
		\left\|
		\tilde{\bf{u}}(\bm{\mu}_i)-\mathbb{V}_k\hat{\textbf{u}}_{N}(\bm{\mu}_i)
		\right\|_{\infty}
	}{
		\left\|
		\tilde{\bf{u}}(\bm{\mu}_i)
		\right\|_{\infty}
	}, \qquad
    	\varepsilon_{max}=  \max\limits_{i\in\{1,\dots,N_t\}}
	\frac{
		\left\|
		\tilde{\bf{u}}(\bm{\mu}_i)-\mathbb{V}_k\hat{\textbf{u}}_{N}(\bm{\mu}_i)
		\right\|_{\infty}
	}{
		\left\|
		\tilde{\bf{u}}(\bm{\mu}_i)
		\right\|_{\infty}
	}.
\end{equation}
where the index $k$ is obtained by evaluating Eq.~\eqref{eqC} and $\hat{\textbf{u}}_{N}(\bm{\mu})$ is the prediction of the reduced solution after evaluating Eq.~\eqref{eq60}. Moreover, the relative average projection error $\varepsilon_{rel}^{proj} $ and relative maximum projection error $\varepsilon_{max}^{proj}$ is defined as:
\begin{equation}
	\varepsilon_{rel}^{proj}=\frac{1}{N_t}\sum_{i=1}^{N_t}
	\frac{
		\left\|
		\tilde{\bf{u}}(\bm{\mu}_i)-\mathbb{V}_k\mathbb{V}_k^T\tilde{\bf{u}}(\bm{\mu}_i)
		\right\|_{\infty}
	}{
		\left\|
		\tilde{\bf{u}}(\bm{\mu}_i)
		\right\|_{\infty}
	}, \qquad
    	\varepsilon_{max}^{proj}=  \max\limits_{i\in\{1,\dots,N_t\}}
	\frac{
		\left\|
		\tilde{\bf{u}}(\bm{\mu}_i)-\mathbb{V}_k\mathbb{V}_k^T\tilde{\bf{u}}(\bm{\mu}_i)
		\right\|_{\infty}
	}{
		\left\|
		\tilde{\bf{u}}(\bm{\mu}_i)
		\right\|_{\infty}
	},
\end{equation} to measure the error stemming from the approximation of the full-order solution with its projection onto the reduced basis space.

\subsection{2D-Plate with edge crack}\label{sec:ex1}
\begin{figure}[!h]
	\centering
	\includegraphics[width=0.5\textwidth]{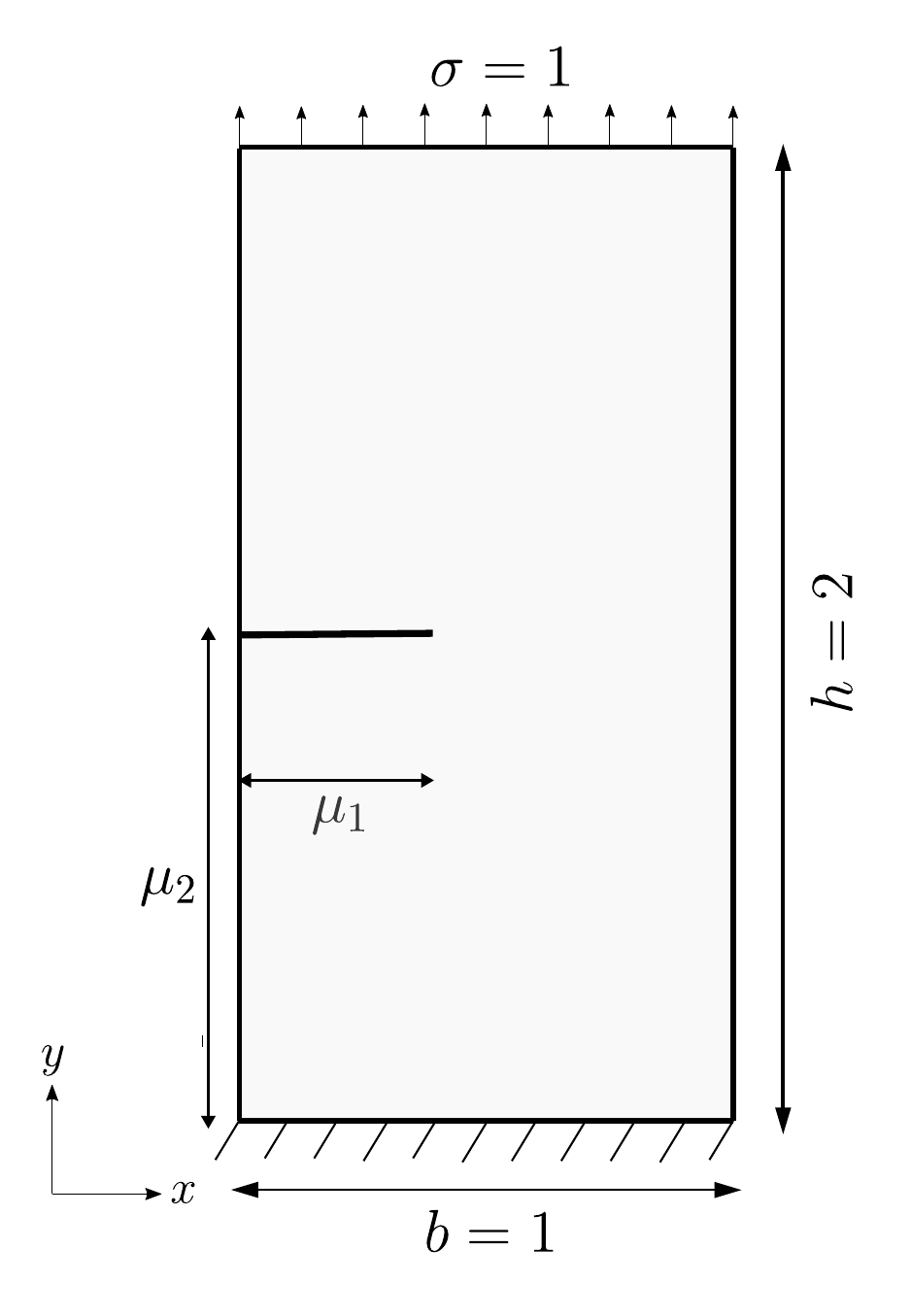} \\
	\caption{Example 5.1: Geometry and parameterization of the plate with edge crack under tension.}\label{fig:2D plate}
\end{figure}

The first numerical example is a 2D plate with edge crack under tension. The geometry and material properties are adopted from the benchmark in \cite{Nguyen2015}. Therefore, the Young's modulus is $E=10^3$ and the Poisson's ratio is $\nu=0.3$ with the assumption of a plain strain condition. The plate is subjected to a tensile stress of $\sigma=1$ along the top edge. Fig.~\ref{fig:2D plate} depicts the geometry and boundary conditions of the problem. Homogeneous Dirichlet boundary conditions are imposed at the bottom edge, that is we fix the displacements in $y$-direction while the bottom left corner is fixed in both $x$- and $y$-direction. In this example, the length and location of the crack is parameterized as depicted in Fig.~\ref{fig:2D plate}. The geometry is discretized with cubic $C^2$-continuous B-splines and a mesh with 1127 elements over a Cartesian grid resulting in $\tilde{\mathcal{N}}_{h}=2400$ degrees of freedom.

\subsubsection{1D geometrical parameterization}\label{sec:ex1_1d}
\begin{figure}[!h]
	\begin{subfigure}[b]{0.49\textwidth}
		\centering
		\includegraphics[width=\textwidth]{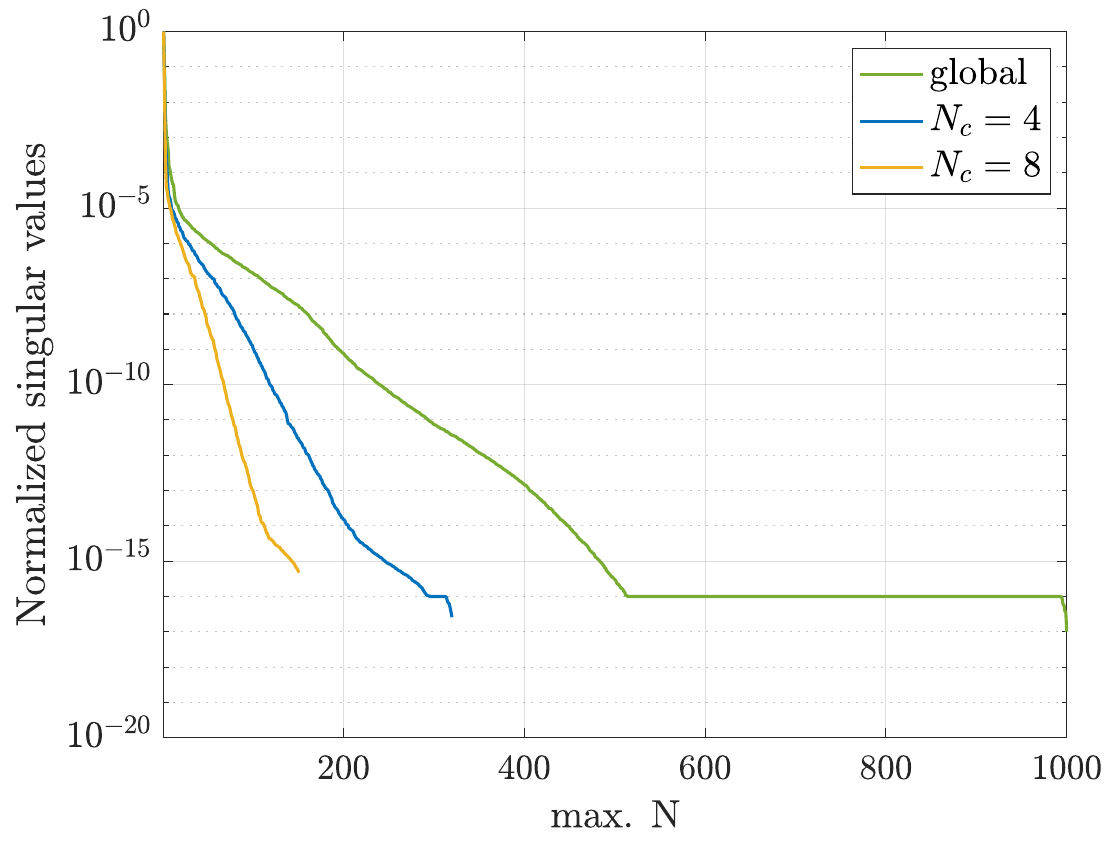}
		\caption{Singular values decay}
		\label{fig:POD_length}
	\end{subfigure}
	\begin{subfigure}[b]{0.49\textwidth}
		\centering
		\includegraphics[width=\textwidth]{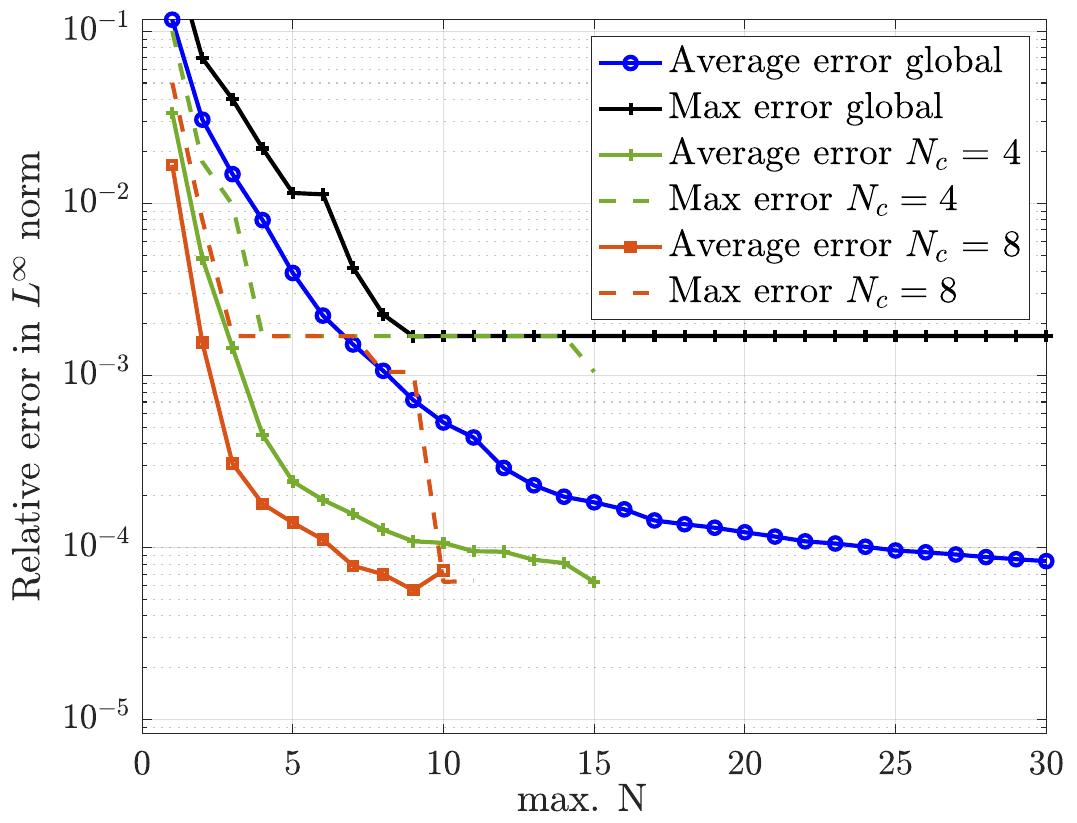}
		\caption{Error analysis}
		\label{fig:error_length}
	\end{subfigure}
	\caption{Example 5.1.1: Singular values decay and relative error in $L^{\infty}$ norm vs.\ maximum number of reduced basis functions $N$ over all the clusters, for different numbers of clusters.}
	\label{fig:results_length}
\end{figure}

First, we consider the problem with one geometrical parameter. The parameter $\mu_1 \in [0.3,0.5]$ represents the length of the crack as depicted in Fig.~\ref{fig:2D plate}. The location of the crack is fixed at the middle of the left edge, that is $\mu_2=1.0$. For this parameterization, we construct a ROM with the localized non-intrusive reduced basis method presented in Sec.~\ref{sec4}. We employ a training set of dimension $N_s=1000$ to construct the reduced basis. Note that this refers to the global dimension and is chosen such that a sufficiently large number of samples is included in each cluster after application of the localization strategy as referred to in Sec.~\ref{sec:cluster}. Fig.~\ref{fig:results_length} depicts the decay of the singular values of the POD for the cluster with the highest number of basis functions and the error analysis for different numbers of clusters. We remark that a test set of dimension $N_t=100$ is employed for the error analysis. It can be observed that increasing the number of clusters allows for a more rapid decay of the singular values and reduces further the dimension of the reduced basis. Note that for the single parameter case where only the length of the crack is parameterized, we employ RBFs to evaluate the error in Fig.~\ref{fig:error_length} as elaborated in Sec.~\ref{sec:localRB}.


\begin{figure}[!h]
	\begin{subfigure}[b]{0.49\textwidth}
		\centering
		\includegraphics[width=\textwidth]{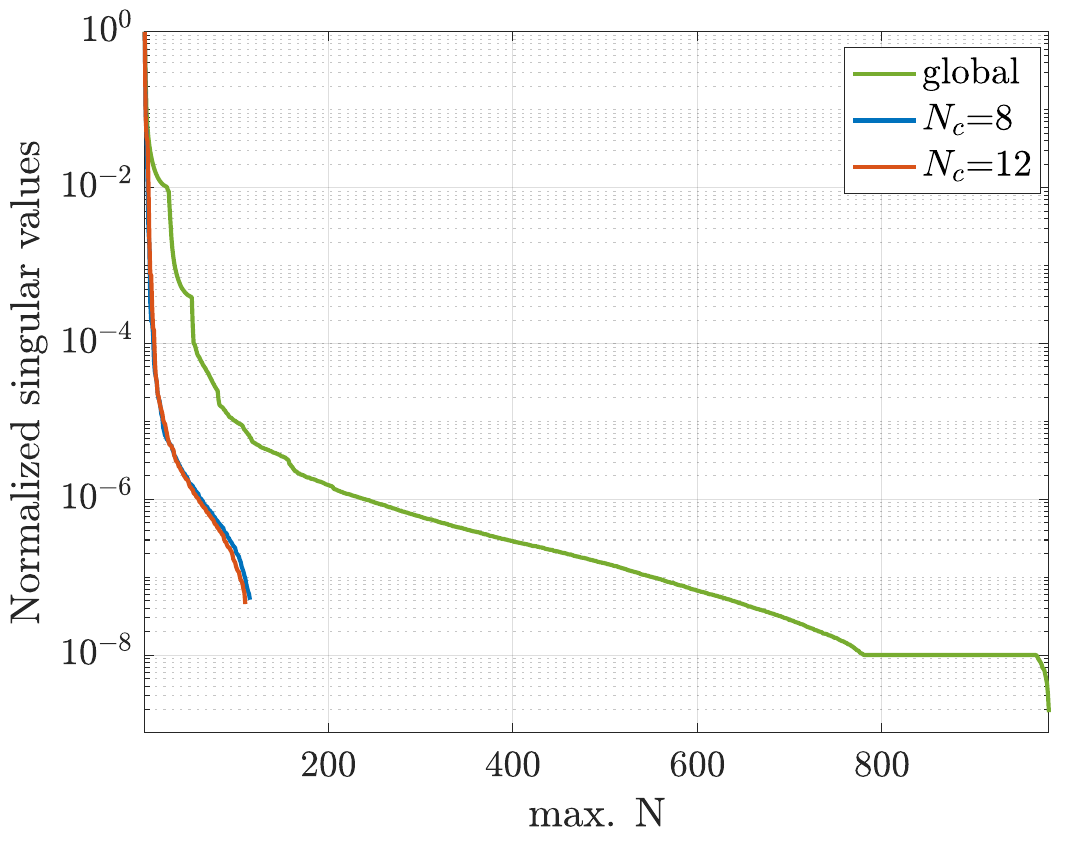}
		\caption{Singular values decay}
		\label{fig:POD_location}
	\end{subfigure}
	\begin{subfigure}[b]{0.49\textwidth}
		\centering
		\includegraphics[width=\textwidth]{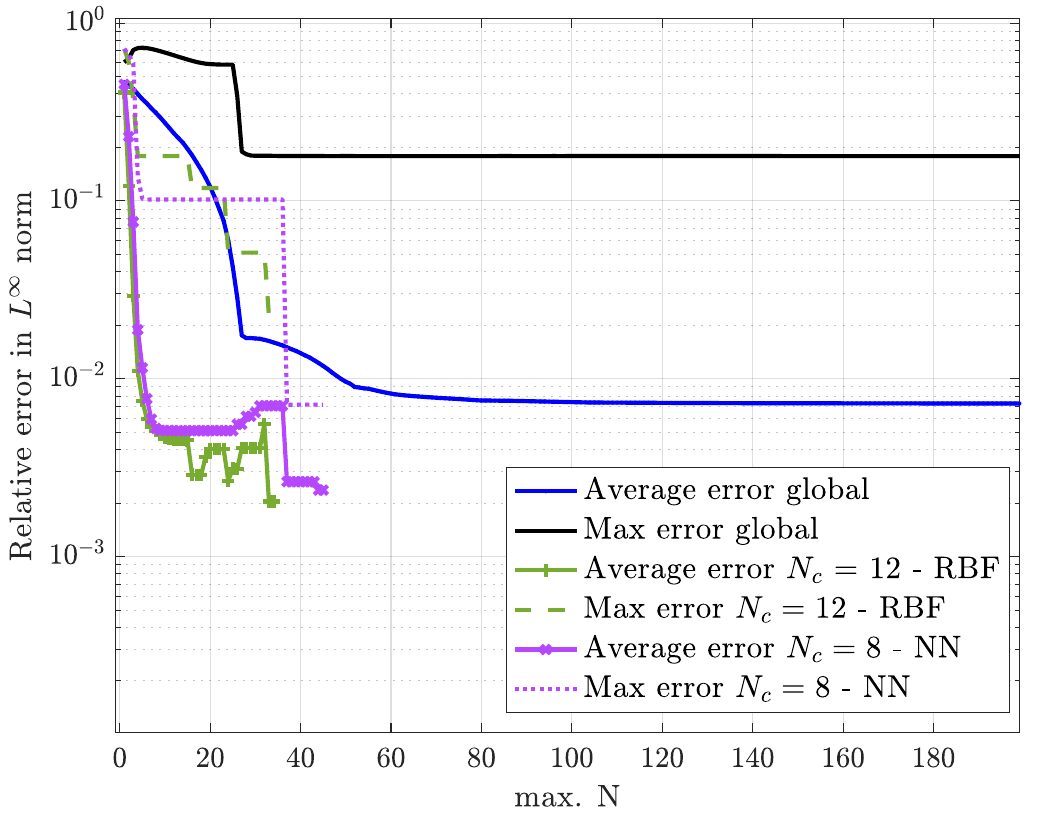}
		\caption{Error analysis}
		\label{fig:error_location}
	\end{subfigure}
	\caption{Example 5.1.1: Singular values decay and relative error in $L^{\infty}$ norm vs.\ maximum number of reduced basis functions $N$ over all the clusters, for different numbers of clusters.}
	\label{fig:results_location}
\end{figure}

Moreover, we consider the case where only the location of the crack is parameterized. The parameter $\mu_2 \in [0.5,1.5]$ represents the location of the crack as depicted in Fig.~\ref{fig:2D plate}, while the length of the crack is fixed as $\mu_1=0.45$. Let us now construct a localized ROM. For this purpose, we employ a training set of dimension $N_s=1000$ to construct the reduced basis. The decay of the singular values of the POD is depicted in Fig.~\ref{fig:POD_location} for the cluster with the highest number of basis functions. It can be observed that the dimension of the reduced basis becomes higher than in the case of parameterized length in Fig.~\ref{fig:POD_length} and the localization significantly reduces the dimension. Next, we employ a test sample of dimension $N_t=100$ for the error analysis whose results are presented in Fig.~\ref{fig:error_location}. 

We observe that the dimension of the ROM is reduced effectively by increasing the number of clusters, while the relative error in $L^{\infty}$ norm reaches an accuracy of the order $10^{-3}$. For comparison, we employ interpolation with RBFs \cite{Powell1987, Buhmann2000}. The results indicate that the localized ROM based on neural networks requires less clusters to achieve a comparable level of accuracy. We remark that these results are obtained using neural networks with 4 hidden layers and 15 neurons per hidden layer for all clusters, while both input and output data are normalized using the minimum and maximum of the respective values for feature scaling within the interval $[-1,1]$. Note that training is performed without regularization and no significant differencies are observed between training and validation loss. Table~\ref{tab:edge1_times} summarizes the main results and obtained computation times for the ROM with $N_c=8$ clusters. All depicted results are CPU times. The online CPU time for the ROM includes the evaluation of the classifier and the projection of the solution into the full space, while the FOM solution time refers to the time required for level set computation, assembly and solution. It should be remarked that the majority of the offline cost is associated to the computation of the snapshots for constructing the reduced basis, while the offline time for training the neural networks is relatively small. Note that each of these tasks can be performed in parallel to further reduce the offline computational cost.

\begin{table}[!h]
	\caption{Example 5.1.1: Number of basis functions  and computational cost.}\label{tab:edge1_times}
	\centering
		\begin{tabular*}{0.65\textwidth}{@{\extracolsep\fill}lc}
			\toprule
			$N_c$   & 8  \\
	     	$\text{min.}$ $N$   & 25  \\
	    	$\text{max.}$ $N$   & 45  \\
           Offline CPU time (data generation) [h]   &  $\approx$ 2.5 \\
	    	Offline CPU time (NN-training) [min]   & 3.63 \\ 
            Average online CPU time [s]   &  0.0128 \\ 
            Standard deviation online CPU time [s] & 0.0065 \\ 
	    	FOM solution time [s]  & 10 \\
            Average solution speedup &  $781\times$ \\ 
			\botrule
		\end{tabular*}
	\end{table}

\subsubsection{2D geometrical parameterization}\label{sec:ex1_2d}
\begin{figure}[!h]
	\begin{subfigure}[b]{0.49\textwidth}
		\centering
		\includegraphics[width=\textwidth]{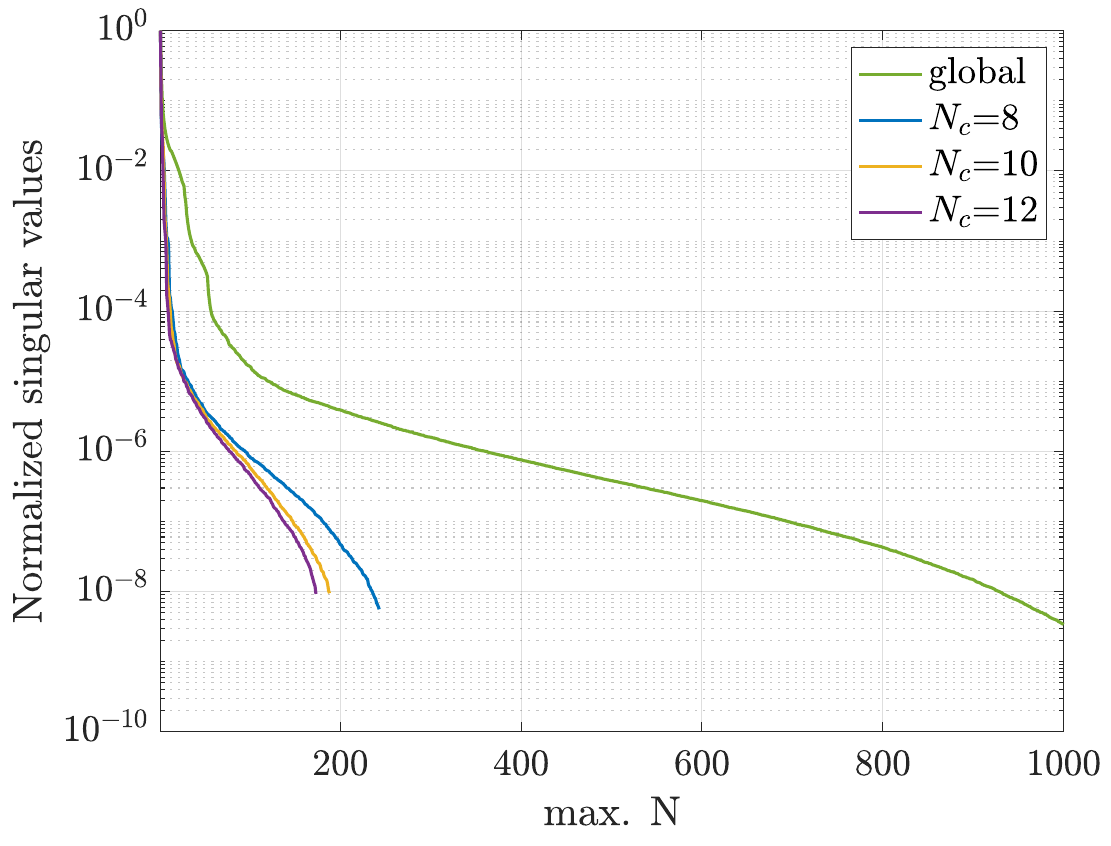}
		\caption{Singular values decay}
		\label{fig:POD_2param}
	\end{subfigure}
	\begin{subfigure}[b]{0.49\textwidth}
		\centering
		\includegraphics[width=\textwidth]{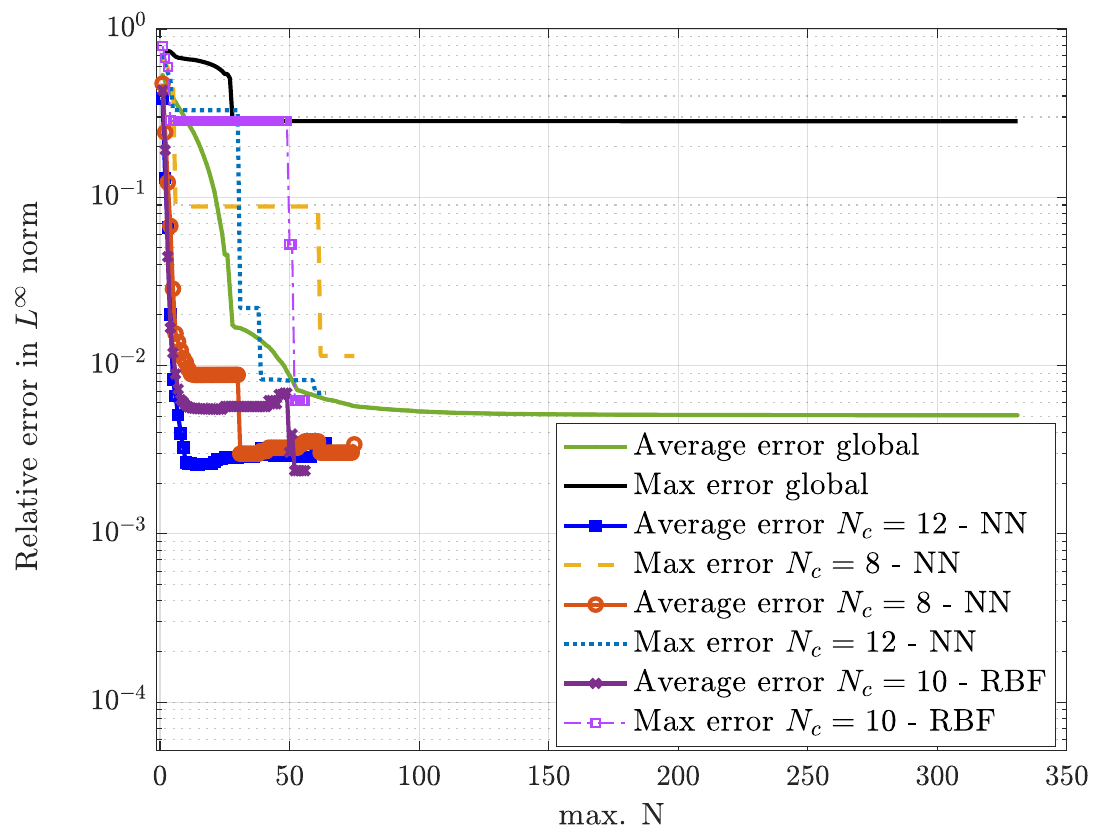}
		\caption{Error analysis}
		\label{fig:error_2param}
	\end{subfigure}
	\caption{Example 5.1.2: Singular values decay and relative error in $L^{\infty}$ norm vs.\ maximum number of reduced basis functions $N$ over all the clusters, for different numbers of clusters.}
	\label{fig:results_2param}
\end{figure}

Next, we consider a two-dimensional geometrical parameterization of the problem, where both the length and location of the crack is parameterized. Here, the parameter vector $\boldsymbol{\mu}=[\mu_1,\mu_2]$ represents the length of the crack $\mu_1$ and the location of the crack $\mu_2$ within the range of $[\mu_1,\mu_2] \in [0.3,0.5] \times [0.5,1.5]$ as depicted in Fig.~\ref{fig:2D plate}. Note that the 2D parameter vector acts as indicator for constructing a classifier in the offline phase as described in Sec.~\ref{sec:classify}. To construct a localized ROM, we employ a training set with a total dimension of $N_s=1500$ samples. The dimension of the training set is chosen as sufficiently large to account for the higher-dimensional parameterization and the complexity of the solution manifold. Fig.~\ref{fig:POD_2param} depicts the decay of the singular values of the POD for different numbers of clusters, while considering for each case the cluster with the highest number of basis functions. Similarly to the previous test case, the dimension of the reduced basis is high and the singular values decay much faster with the localization. Furthermore, we employ a test sample of dimension $N_t=100$ for the error analysis, which yields the results depicted in Fig.~\ref{fig:error_2param}. It can be observed that increasing the number of clusters leads to more accurate results while it further reduces the dimension of the reduced basis. Note that the local ROM with $N_c=12$ clusters is more accurate than the one with $N_c=8$ clusters for a dimension of the reduced basis up to $N \approx 30$, while the error results are almost identical for increasing dimension of the reduced basis. We remark that the same network architecture is employed for all clusters as in Sec.~\ref{sec:ex1_1d}, while the input/output data are normalized according to Sec.~\ref{sec:localRB}. The number of neurons per hidenn layer is the same for all clusters, while some of the clusters contain a smaller amount of data than others. To prevent overfitting, we employ $L^2$ weight regularization and set the parameter that controls regularization to $\lambda=10^{-4}$. Note that although the chosen feature scaling and regularization can affect the speed of the training process, our numerical investigations show that it yields the most accurate results. Table~\ref{tab:edge2_times} summarizes the main results and obtained computation times for the ROM with $N_c=8$ clusters.

\begin{table}[!h]
	\caption{Example 5.1.2: Number of basis functions  and computational cost.}\label{tab:edge2_times}
	\centering
		\begin{tabular*}{0.65\textwidth}{@{\extracolsep\fill}lc}
			\toprule
			$N_c$   & 8  \\
			$\text{min.}$ $N$   & 30  \\ 
			$\text{max.}$ $N$   & 75  \\ 
            Offline CPU time (data generation) [h]   &  $\approx$ 4 \\
			Offline CPU time (NN-training) [min]   & 40 \\ 
            Average online CPU time [s]   &  0.0144 \\ 
            Standard deviation online CPU time [s] & 0.0125 \\ 
		    FOM solution time [s]  & 10 \\
            Average solution speedup &  $694\times$ \\ 
			\botrule
		\end{tabular*}
	\end{table}

\begin{figure}[!h]
	\begin{subfigure}[b]{1.0\textwidth}
		\centering
		\includegraphics[width=\textwidth]{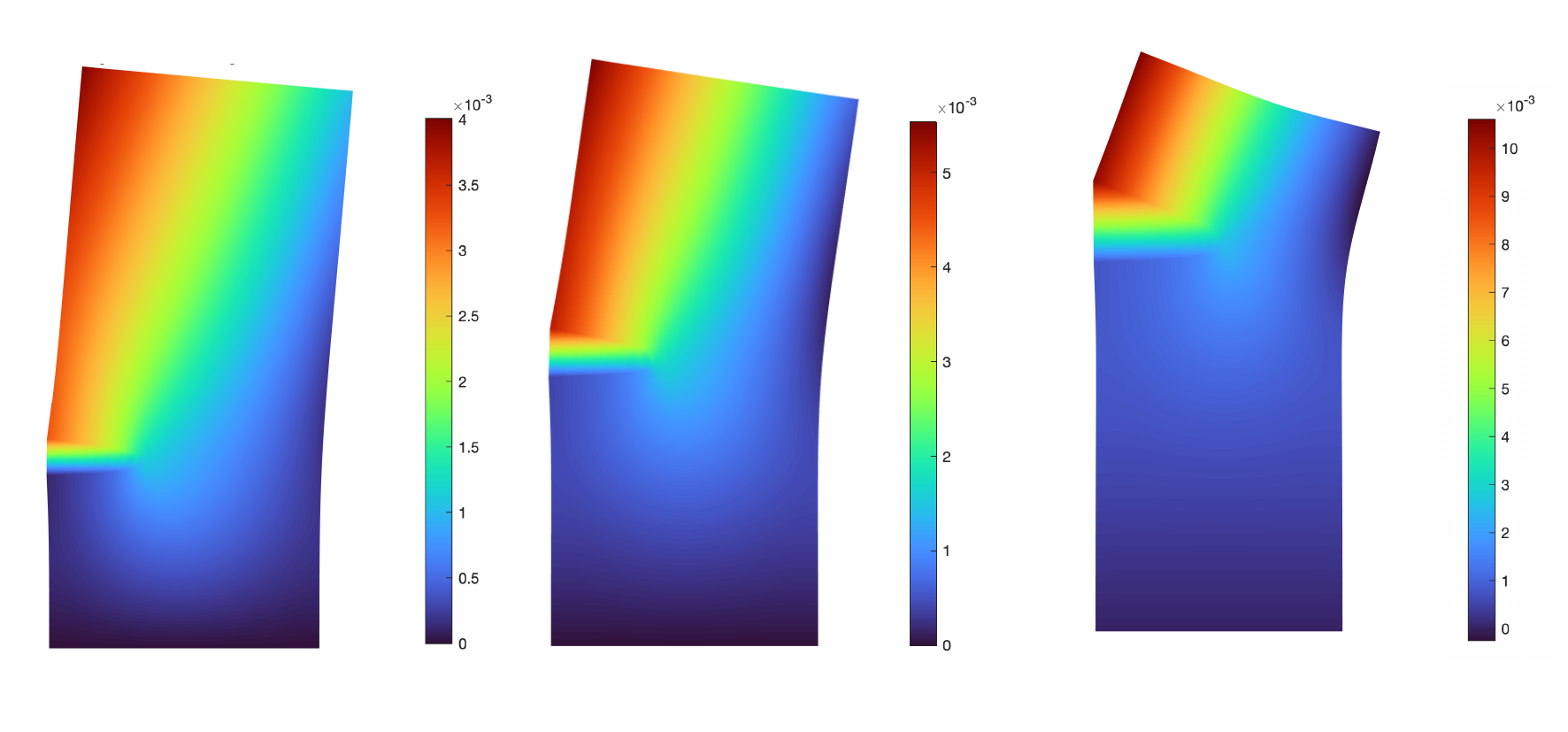}
		\caption{Displacement solutions with FOM}
		\label{fig:FOM_solutions_edge}
	\end{subfigure}
     \hfill
	\begin{subfigure}[b]{1.0\textwidth}
		\centering
		\includegraphics[width=\textwidth]{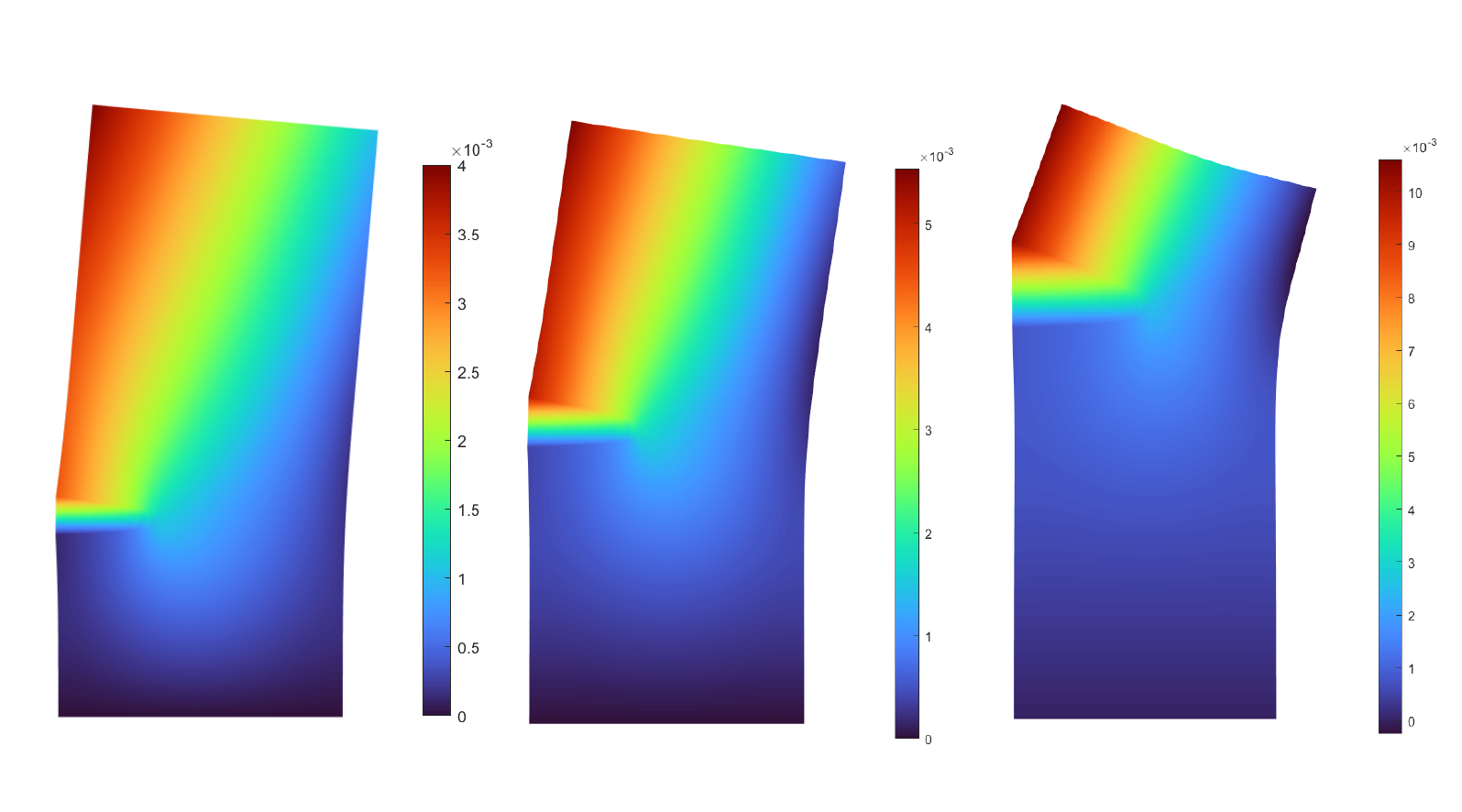}
		\caption{Displacement solutions with ROM}
		\label{fig:ROM_solutions_edge}
	\end{subfigure}
	\caption{Example 5.1.2: Vertical displacement solutions computed with the FOM (a) and ROM (b) for three parameter values  $\mu_1=[0.329, 0.398, 0.494]$ and $\mu_2=[0.649, 0.990, 1.473]$. Deformed configuration enlarged by a factor of 30.}
	\label{fig:edge_FOM-ROM}
\end{figure}
\begin{figure}[!h]
	\centering
	\includegraphics[width=1.0\textwidth]{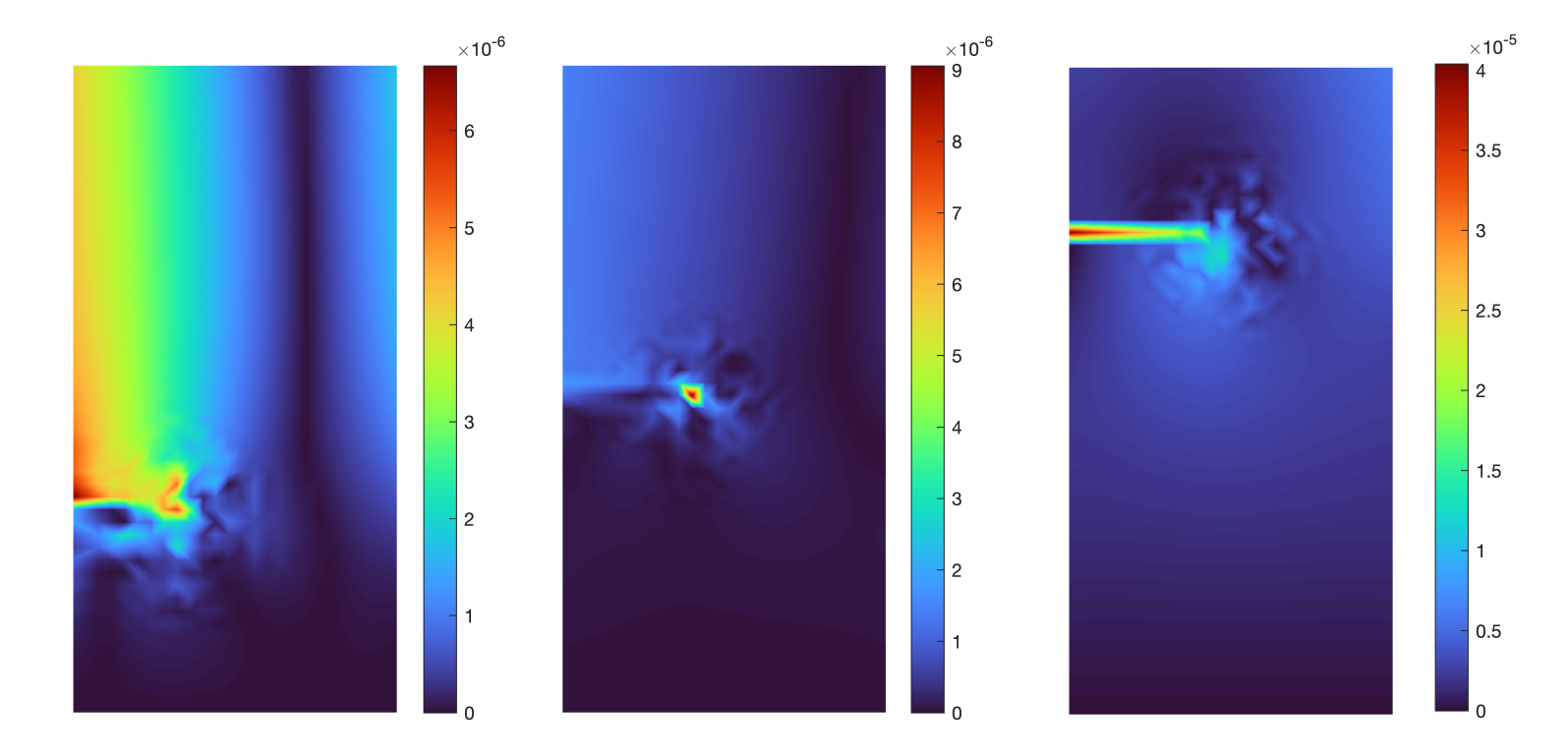} \\
	\caption{Example 5.1.2: Point-wise absolute errors between the vertical displacement solutions computed with the FOM and ROM for three parameter values $\mu_1=[0.329, 0.398, 0.494]$ and $\mu_2=[0.649, 0.990, 1.473]$.}\label{fig:diff}
\end{figure}

Finally, Fig.~\ref{fig:edge_FOM-ROM} depicts the solutions of the local ROM for three different values of the test sample in comparison with the respective solutions of the FOM on the deformed configuration. Note that an enlargement factor is applied for visualization. It can be observed that despite the strong variation in the solution for different values of the parameters, a good qualitative agreement can be achieved with the localization. Fig.~\ref{fig:diff} shows the corresponding point-wise absolute errors between the FOM and ROM, which are in the order of $10^{-5}$.

\subsection{2D-Plate with center crack}\label{sec:ex2}
\begin{figure}[!h]
	\centering
	\includegraphics[width=0.6\textwidth]{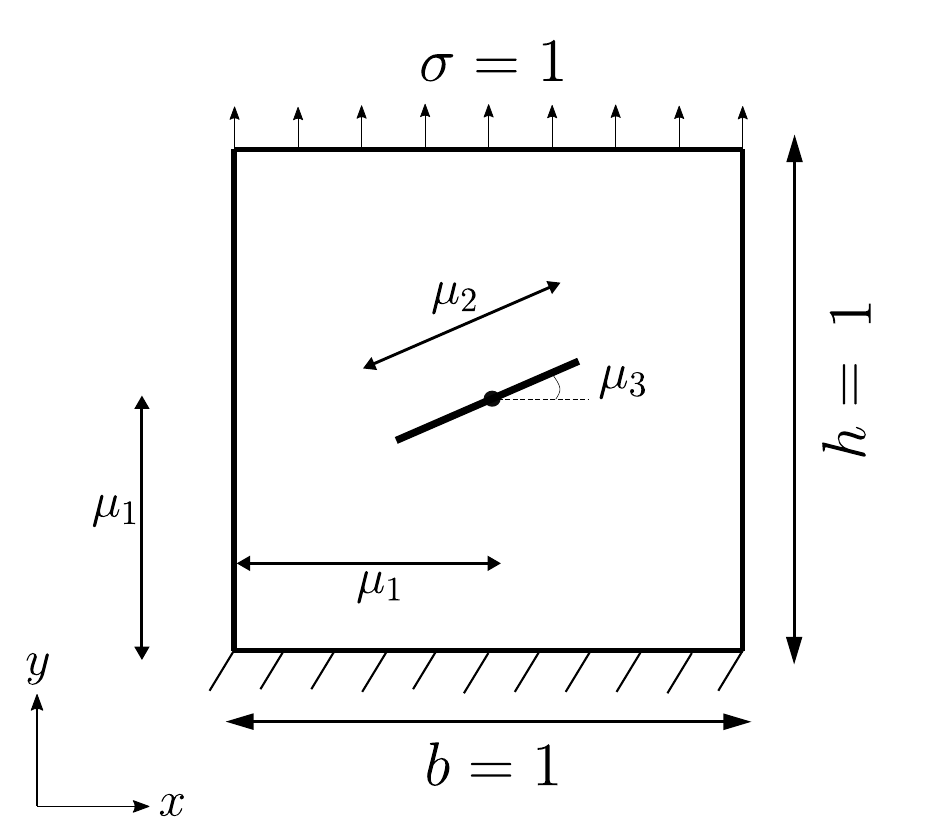} \\
	\caption{Example 5.2: Geometry and parameterization of the plate with center crack under tension.}\label{fig:2D plate_center}
\end{figure}

This example investigates the capabilities of the ROM approach for problems with multiple geometric parameters. For this purpose we employ a 2D plate with center crack under tension. The geometry and material properties are adopted from the benchmark in \cite{Agathos2020}. Thus, the Young's modulus is $E=100$ and the Poisson's ratio is $\nu=0.3$ under plane strain conditions. The plate is subjected to a tensile stress of $\sigma=1$ along the top edge. Fig.~\ref{fig:2D plate_center} depicts the geometry and boundary conditions of the problem. We impose homogeneous Dirichlet boundary conditions along the bottom edge, that is we fix the displacements in $y$-direction while the bottom left corner is fixed in both $x$- and $y$-direction. The parameterization of the problem is depicted in Fig.~\ref{fig:2D plate_center}. We consider three parameters with the parameter vector $\mu=[\mu_1,\mu_2,\mu_3]$, where the parameter $\mu_1 \in [0.2,0.7]$ represents the coordinates of the center of the crack, $\mu_2 \in [0.15,0.2]$ represents the length of the crack and $\mu_3 \in [-\frac{\pi}{4},\frac{\pi}{4}]$ its orientation as shown in Fig.~\ref{fig:2D plate_center}. We remark that the crack is centered at $(\mu_1,\mu_1)=(x_c,y_c)$ and thus moves along one diagonal of the plate. The geometry is discretized with $C^2$-continuous B-splines, which yields a total of $\tilde{\mathcal{N}}_{h}=2888$ degrees of freedom. 

\begin{figure}[!h]
	\begin{subfigure}[b]{0.49\textwidth}
		\centering
		\includegraphics[width=\textwidth]{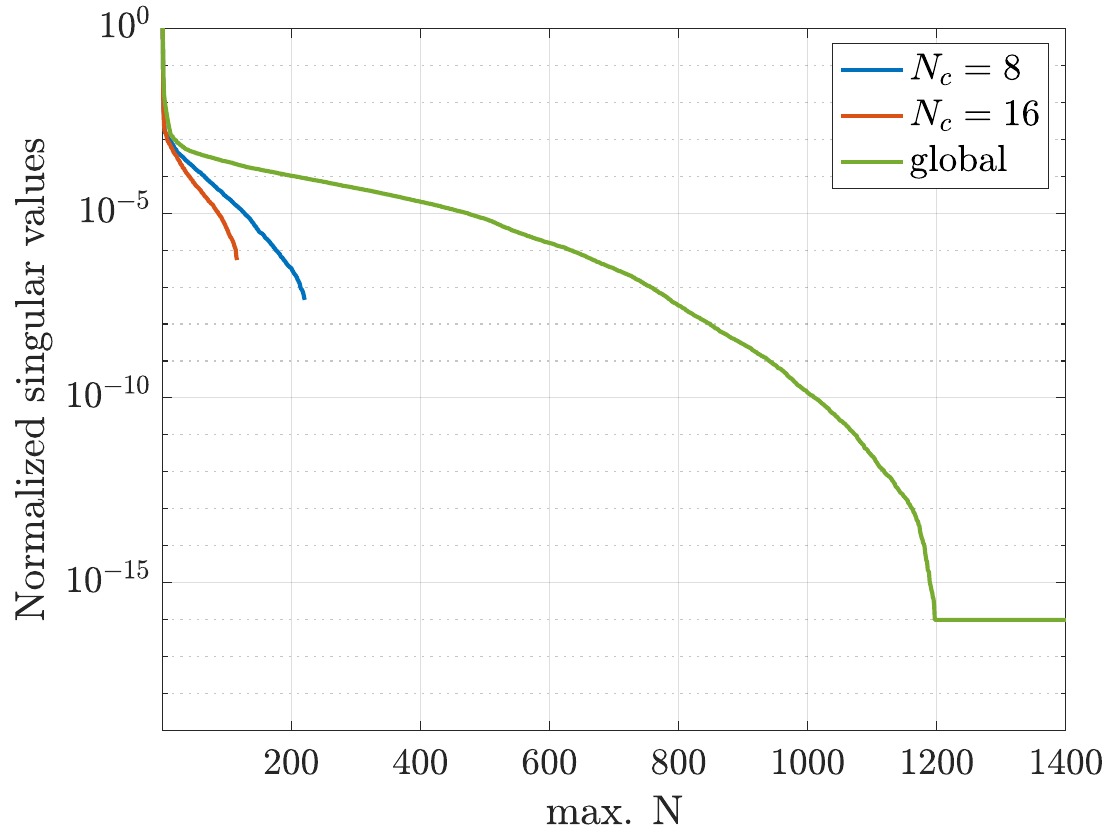}
		\caption{Singular values decay}
		\label{fig:POD_center}
	\end{subfigure}
	\begin{subfigure}[b]{0.49\textwidth}
		\centering
		\includegraphics[width=\textwidth]{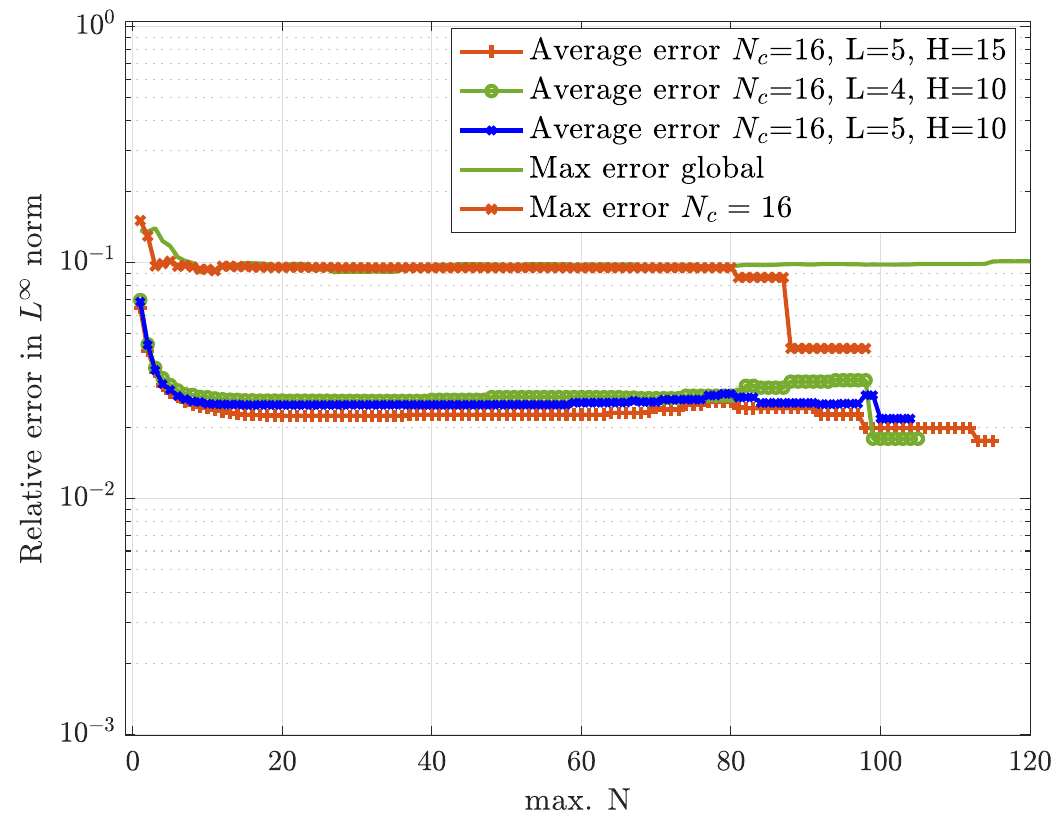}
		\caption{Error analysis}
		\label{fig:error_center}
	\end{subfigure}
	\caption{Example 5.2: Singular values decay for different numbers of clusters and relative error in $L^{\infty}$ norm vs.\ maximum number of reduced basis functions $N$ over all the clusters for different network configurations.}
	\label{fig:results_center}
\end{figure}

Next, we construct a ROM for our problem. For this purpose, we employ a training sample of dimension $N_s=1500$ and construct a global reduced basis as a first step. Fig.\ref{fig:POD_center} shows the decay of the singular values of the POD for different numbers of clusters. Note that the depicted curves refer to the cluster with the highest number of reduced basis functions. It can be observed that the dimension of the global ROM is higher and the decay is slower than the previous test cases. This is due to the fact that the parameterization is more complex, that is the solution manifold becomes more difficult to approximate with one global reduced basis. Nevertheless, localization renders our problem reducible. Furthermore, we generate a test sample of dimension $N_t=100$ to perform the error analysis, whose results are illustrated in Fig.\ref{fig:error_center} for different network architectures. 

\begin{figure}[!h]
	\begin{subfigure}[b]{1.0\textwidth}
		\centering
		\includegraphics[width=\textwidth]{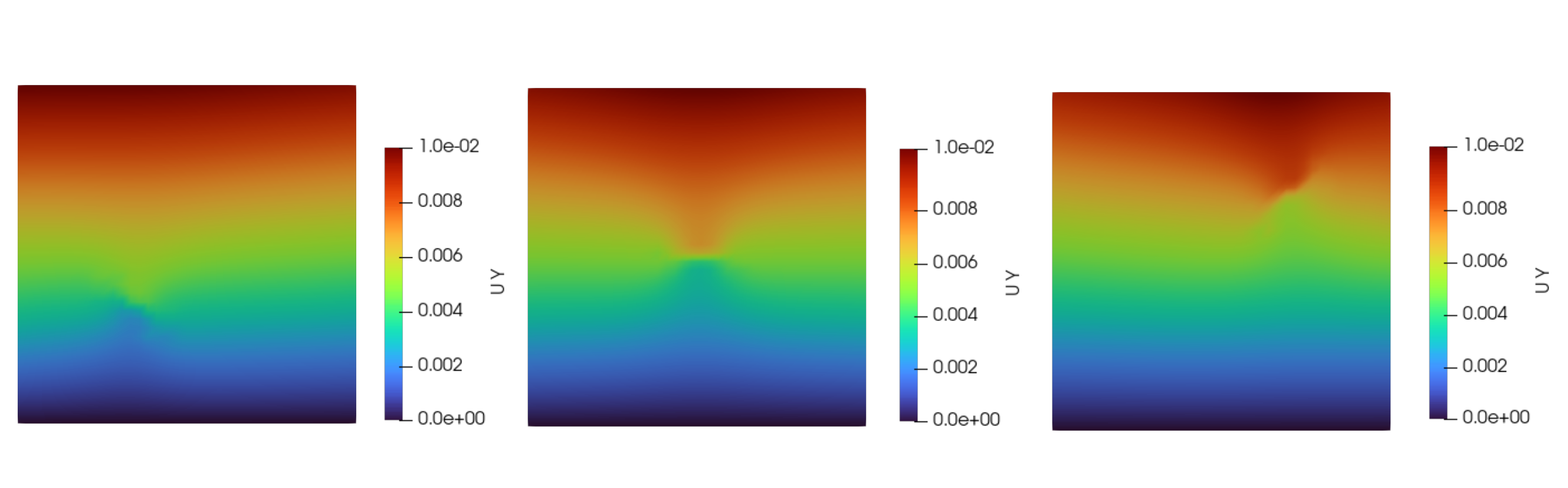}
		\caption{Displacement solutions with FOM}
		\label{fig:FOM_solutions_center}
	\end{subfigure}
	\hfill
	\begin{subfigure}[b]{1.0\textwidth}
		\centering
		\includegraphics[width=\textwidth]{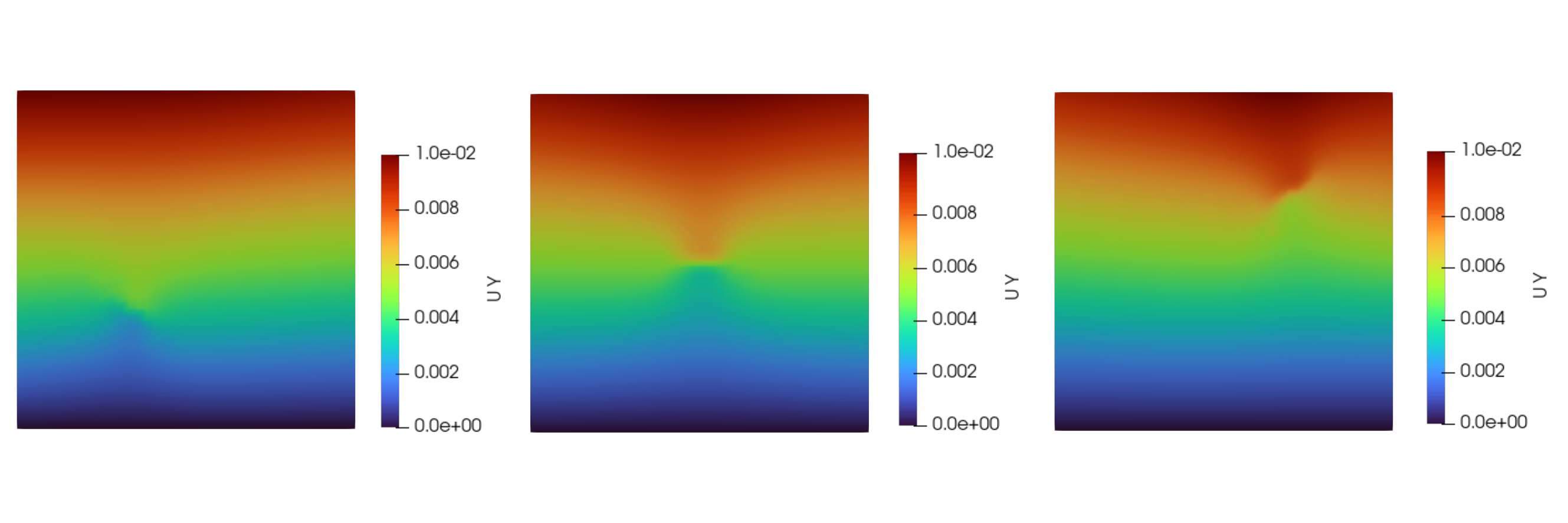}
		\caption{Displacement solutions with ROM}
		\label{fig:ROM_solutions_center}
	\end{subfigure}
    	\begin{subfigure}[b]{1.0\textwidth}
		\centering
		\includegraphics[width=\textwidth]{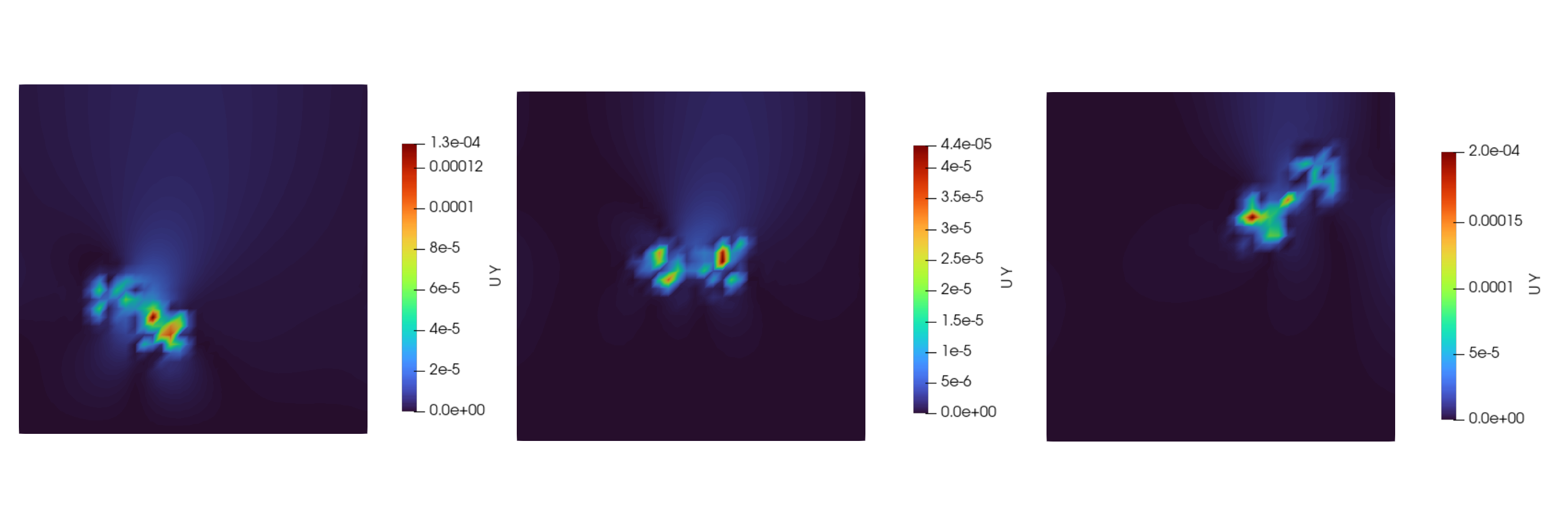}
		\caption{Point-wise absolute errors between FOM and ROM}
		\label{fig:center_diff}
	\end{subfigure}
	\caption{Example 5.2: Vertical displacement solutions computed with the FOM (a), ROM (b) and point-wise errors (c) for three parameter values $\mu_1=[0.344, 0.155, -0.610]$, $\mu_2=[0.508, 0.176, 0.033]$ and $\mu_3=[0.699, 0.199, 0.785]$.}
	\label{fig:center_FOM-ROM}
\end{figure}

We remark that the results are obtained using neural networks with $L=4,5$ hidden layers and $H=10,15$ neurons per layer for all clusters, while Bayesian regularization is employed to prevent overfitting \cite{MacKay1992}. The results indicate that the ROM based on neural networks with 5 hidden layers and 15 neurons per hidden layer is slightly more accurate, while an accuracy of the order $10^{-2}$ is achieved in all cases. Note that further improvement of the convergence behavior is in principle possible, for instance, by augmenting the ROM with physics-informed neural networks \cite{Chen2021}. Table~\ref{tab:center_times} summarizes the main results and obtained computation times for the ROM with $N_c=16$ clusters and the above mentioned network configuration. Moreover, Fig.\ref{fig:center_FOM-ROM} provides a qualitative comparison between the constructed ROM and the FOM for the vertical displacements solutions and three parameters of the test sample as well as the associated point-wise absolute errors. The highest errors are observed around the crack tips and are of the order $10^{-4}$. The comparison reveals that the ROM is able to capture variations in the crack configuration with sufficient accuracy.

\begin{table}[!h]
\caption{Example 5.2: Number of basis functions and computational cost.}\label{tab:center_times}
\centering
	\begin{tabular*}{0.65\textwidth}{@{\extracolsep\fill}lc}
		\toprule
		$N_c$   & 16  \\
		$\text{min.}$ $N$   & 36  \\  
		$\text{max.}$ $N$   & 115  \\ 
	    Offline CPU time (NN-training) [min]   & 63 \\ 
        Average online CPU time [s]   & 0.0156 \\ 
        Standard deviation online CPU time [s] &  0.0133 \\ 
		FOM solution time [s]  & 17 \\
        Average solution speedup &  $1089\times$ \\ 
		\botrule
	\end{tabular*}
\end{table}


\subsection{3D-Plate with edge crack}\label{sec:ex3}
\begin{figure}[!h]
	\centering
	\includegraphics[width=0.6\textwidth]{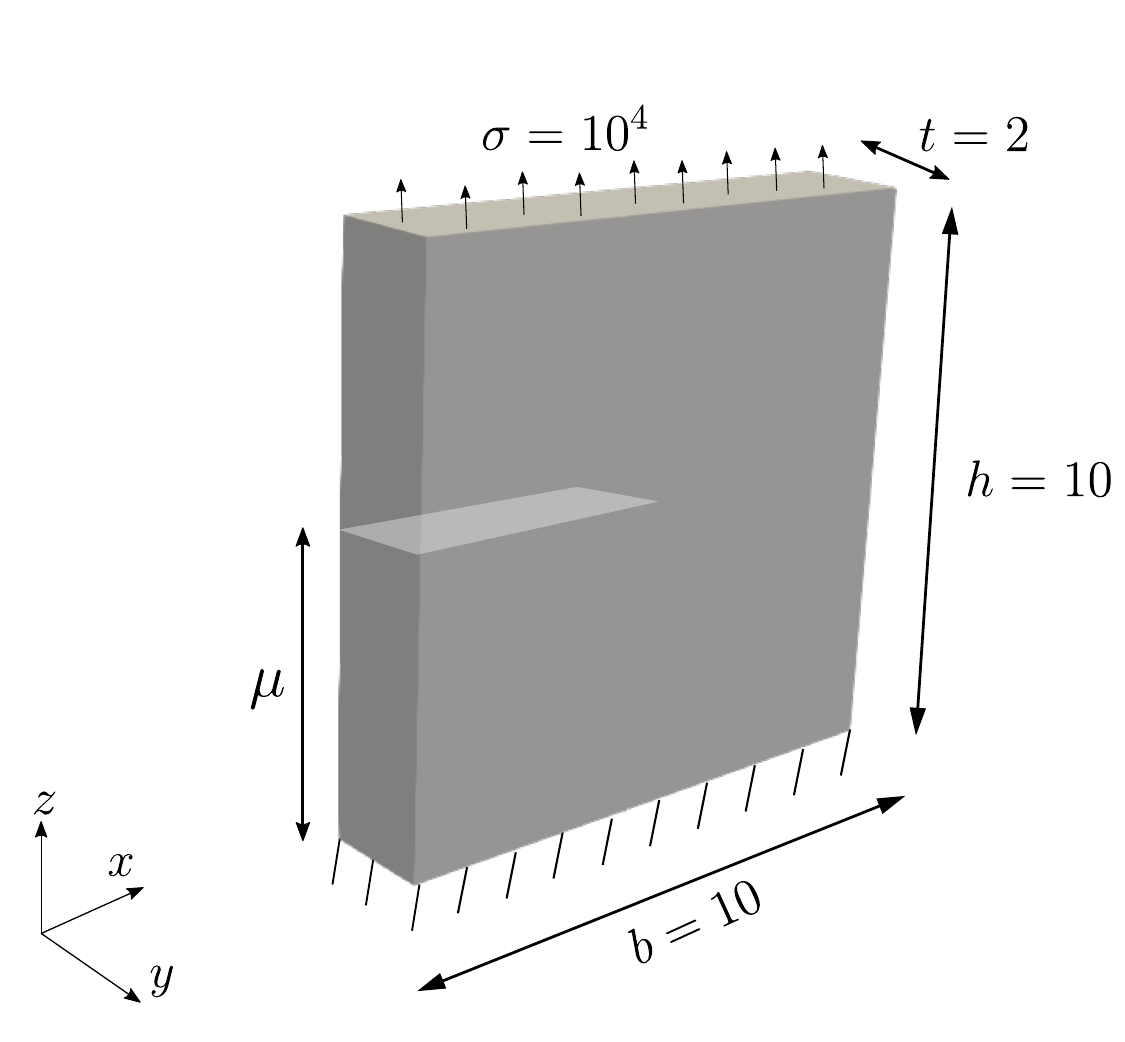} \\
	\caption{Example 5.3: Geometry and parameterization of the 3D plate with edge crack under tension.}\label{fig:3D plate_edge}
\end{figure}

This example aims to show the applicability of the proposed ROM framework to three-dimensional geometries. For this purpose we employ a 3D plate with edge crack under tension. The geometry and material properties are adopted from \cite{Nguyen2015}. The Young's modulus is $E=10^7 \ \text{N/mm}^2$, the Poisson's ratio is $\nu=0.3$ and the prescribed tensile stress is $\sigma=10^4 \ \text{N/mm}^2$. The latter is applied along the top surface of the plate similarly to the example in Sec.~\ref{sec:ex1}. The vertical displacements in all directions are fixed to apply a homogeneous Dirichlet boundary condition along the bottom surface of the plate. The geometry and boundary conditions are illustrated in Fig.~\ref{fig:3D plate_edge}. For our problem we consider one geometrical parameter $\mu \in [2.5,7.5]$ representing the location of the crack as depicted in Fig.~\ref{fig:3D plate_edge}. The length of the crack is set to $a=5$. The geometry is discretized with $C^1$-continuous B-splines and 722 elements over a Cartesian grid, which results in $\tilde{\mathcal{N}}_{h,0}=3600$ degrees of freedom.

\begin{figure}[!h]
	\begin{subfigure}[b]{0.49\textwidth}
		\centering
		\includegraphics[width=\textwidth]{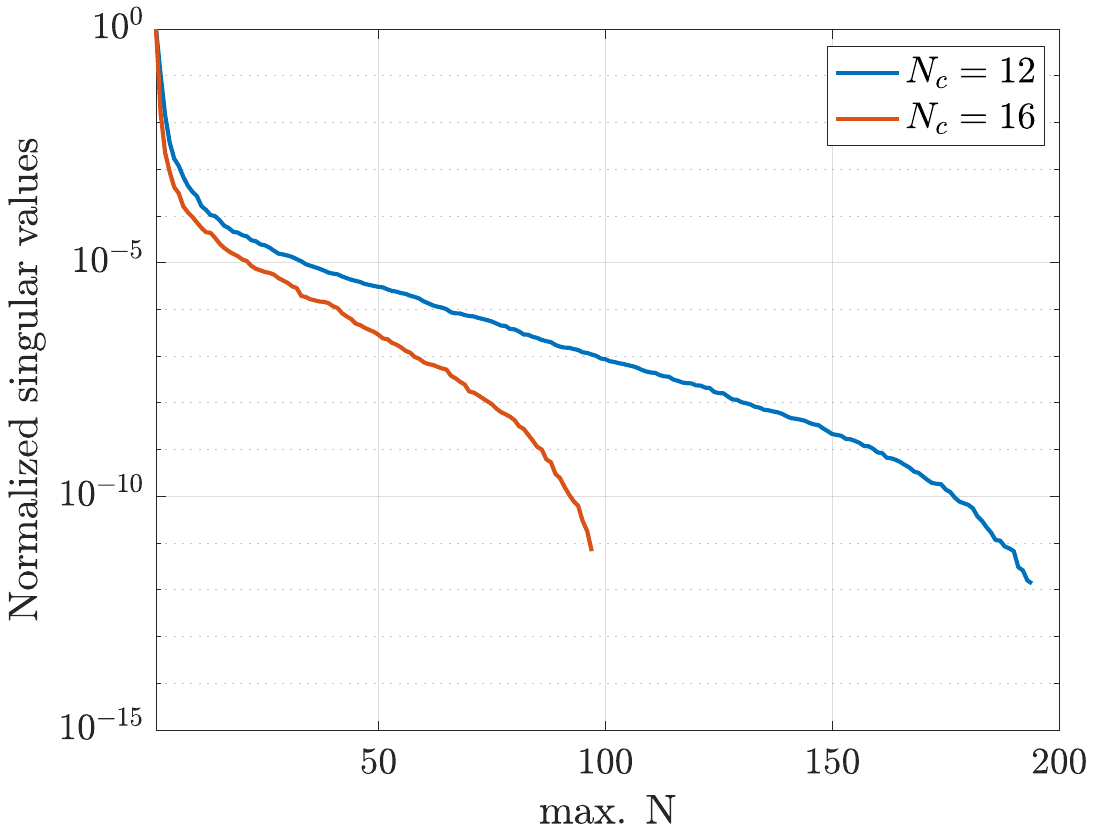}
		\caption{Singular values decay}
		\label{fig:POD_3d}
	\end{subfigure}
	\begin{subfigure}[b]{0.49\textwidth}
		\centering
		\includegraphics[width=\textwidth]{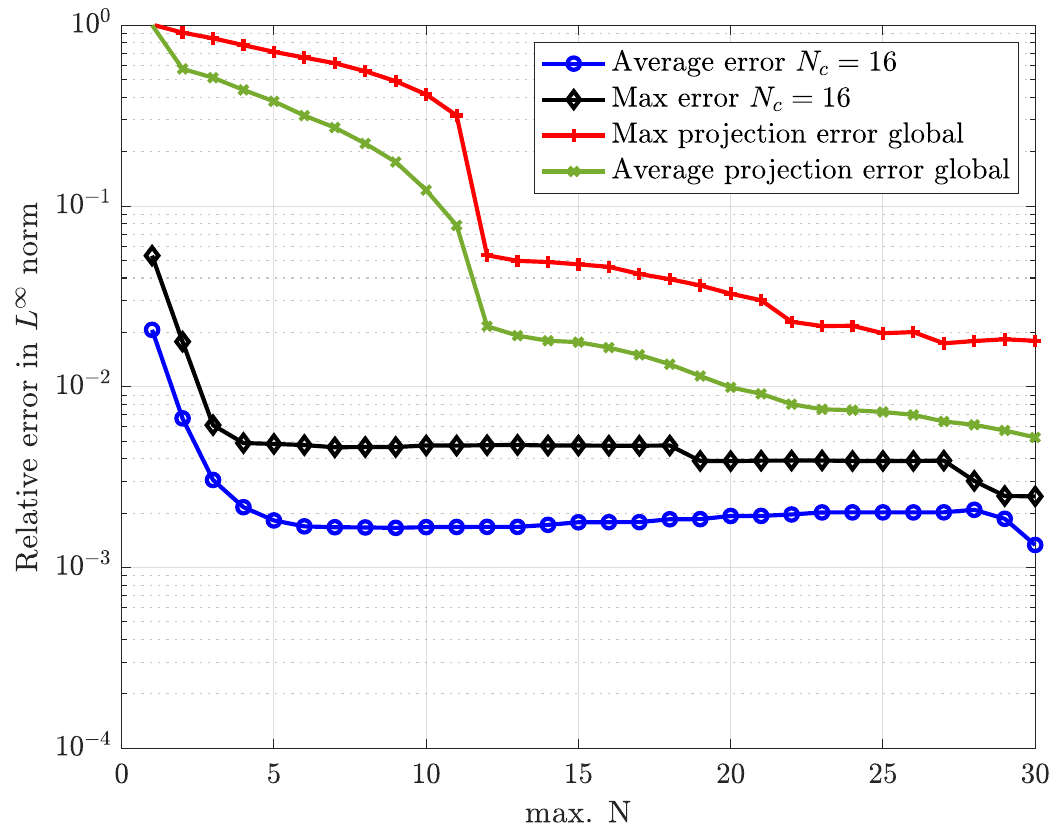}
		\caption{Error analysis}
		\label{fig:error_3d}
	\end{subfigure}
	\caption{Example 5.3: Singular values decay for different numbers of clusters and relative error in $L^{\infty}$ norm vs.\ maximum number of reduced basis functions $N$ over all the clusters for $N_c=16$.}
	\label{fig:results_3d}
\end{figure}

\begin{figure}[!h]
	\begin{subfigure}[b]{1.0\textwidth}
		\centering
		\includegraphics[width=\textwidth]{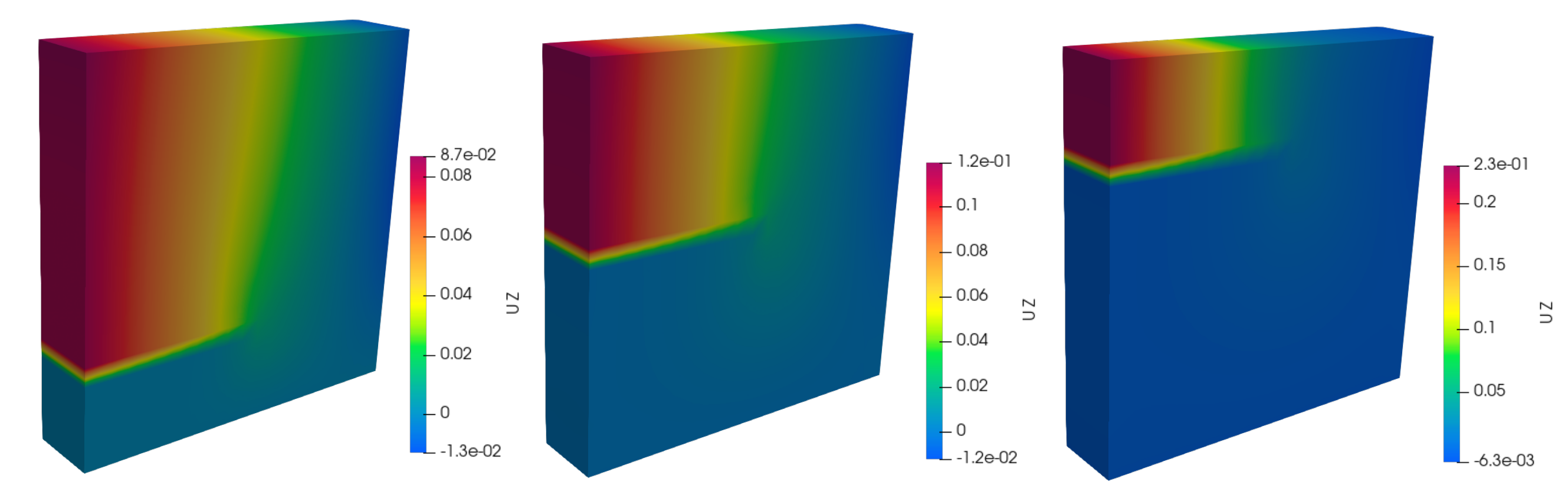}
		\caption{Displacement solutions with FOM}
		\label{fig:FOM_solutions_3d}
	\end{subfigure}
	\hfill
	\begin{subfigure}[b]{1.0\textwidth}
		\centering
		\includegraphics[width=\textwidth]{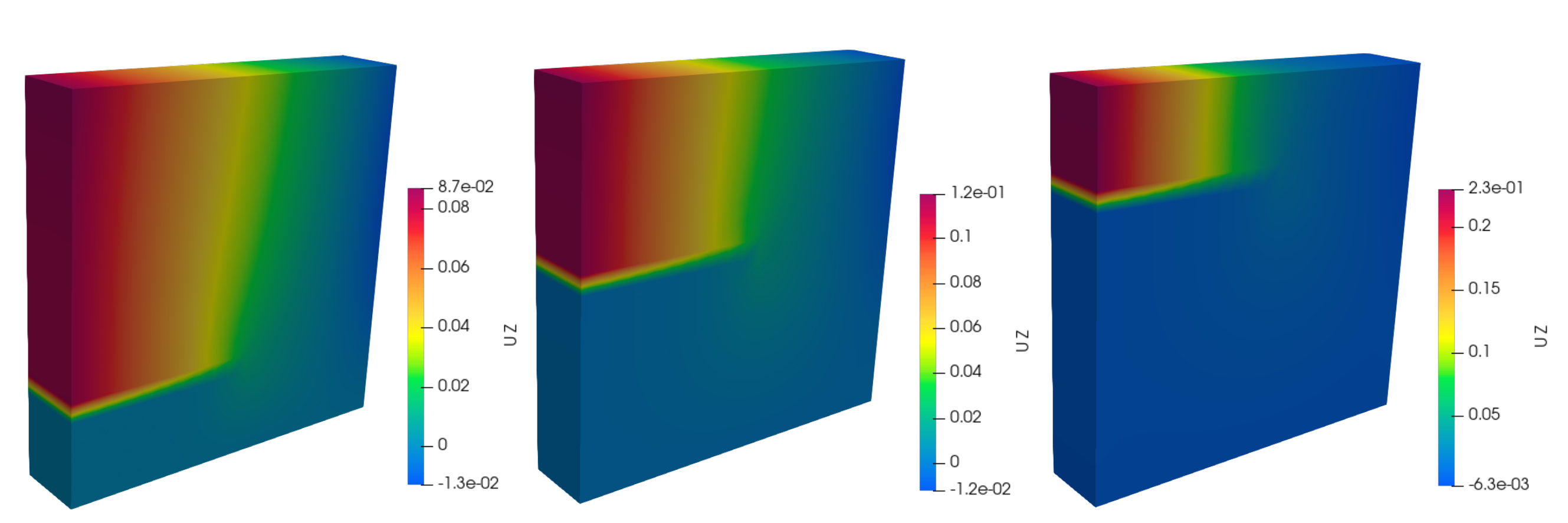}
		\caption{Displacement solutions with ROM}
		\label{fig:ROM_solutions_3d}
	\end{subfigure}
    	\begin{subfigure}[b]{1.0\textwidth}
		\centering
		\includegraphics[width=\textwidth]{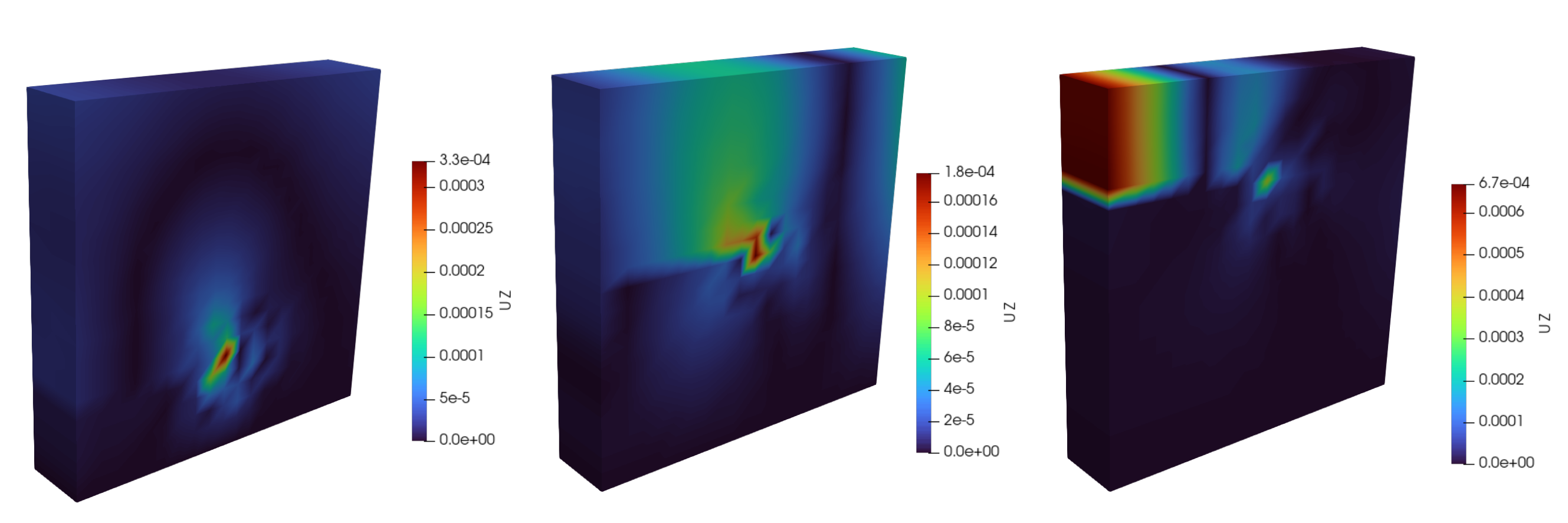}
		\caption{Point-wise absolute errors between FOM and ROM}
		\label{fig:diff_3d}
	\end{subfigure}
	\caption{Example 5.3: Vertical displacement solutions computed with the FOM (a),  ROM (b) and point-wise errors (c) for three parameter values $\mu= [0.251, 0.529, 0.745]$.}
	\label{fig:3d_FOM-ROM}
\end{figure}

Next, we construct a ROM for our problem based on the localization strategy described in Sec.~\ref{sec4}. For this purpose we employ a training sample of dimension $N_{s}=1000$. Fig.~\ref{fig:results_3d} depicts the decay of the singular values of the POD for the respective clusters with the highest number of reduced basis functions as well as the relative error for $N_c=16$. We remark that the latter is performed using a test set of dimension $N_t=100$. The results indicate that localization allows to effectively reduce the dimension of the reduced basis, while an accuracy of $10^{-3}$ is achieved in the $L^{\infty}$ norm. Note that these results are obtained using neural networks with 4 layers and 15 neurons per hidden layer for all clusters, while Bayesian regularization is employed to prevent overfitting. In addition, the projection error of the global ROM in depicted for comparison in Fig.~\ref{fig:error_3d}. We observe that the projection error is higher than the error of the local ROM, indicating poor approximation properties of the global reduced basis. Moreover, Table~\ref{tab:3d_times} summarizes the main computation times and obtained results. Finally, a qualitative comparison between the localized ROM with $N_c=16$ clusters and the FOM is provided in Fig.~\ref{fig:3d_FOM-ROM} for three different parameters of the test sample. Fig.~\ref{fig:diff_3d} depicts the corresponding point-wise errors between the FOM and ROM with an accuracy in the order of $10^{-4}$.

\begin{table}[!h]
\caption{Example 5.3: Number of basis functions and computational cost.}\label{tab:3d_times}
\centering
	\begin{tabular*}{0.65\textwidth}{@{\extracolsep\fill}lc}
		\toprule
		$N_c$   & 16  \\
		$\text{min.}$ $N$   & 13  \\
		$\text{max.}$ $N$   & 30  \\
		Offline CPU time (NN-training) [min]   & 4 \\ 
        Average online CPU time [s]   &  0.0156 \\ 
        Standard deviation online CPU time [s] &  0.0167 \\ 
		FOM solution time [s]  & 35 \\
        Average solution speedup & $2243\times$ \\ 
		\botrule
	\end{tabular*}
\end{table}


\section{Conclusion}\label{sec6}
This work presents a model reduction strategy for parametric problems formulated on domains with cracks and discretized by splines. The proposed strategy targets parametrized problems with geometric parameters that describe the configuration of the crack, such as its size and location. To achieve efficient reduction, localization is employed in the context of non-intrusive reduced basis methods. We obtain local ROMs of small dimension, thus allowing an efficient offline/online decomposition with low online cost.

We have investigated numerically the proposed strategy through several test problems. For this purpose, we considered spline discretizations with XIGA to model the crack. The framework was demonstrated on parameterized problems with multiple parameters describing the crack including one 3D geometry. The results we obtained with the localization strategy indicate the efficiency of the constructed non-intrusive ROMs and high speedup compared to the FOM. Moreover, the local ROMs were more accurate than their global counterparts, while requiring less reduced basis functions to achieve the same level of accuracy. 
To sum up with, the proposed framework allows efficient reduction for geometrically parameterized problems thus allowing to explore different crack configurations in a many-query and real-time context.

Future research involves the integration of the ROM framework in the context of inverse problems to facilitate damage detection tasks. Furthermore, this work focuses on linear elastic problems with geometric parameters describing the configuration of the crack. The extension to more complex scenarios, such as dynamic problems, nonlinear behavior, crack propagation etc., is a subject of future work. From the model reduction point of view, nonlinear dimensionality reduction techniques may be beneficial for scenarios with increasing complexity of the underlying physical problem at hand.

\section*{Acknowledgments}\label{sec7}
 The financial support under the Excellence Strategy of the Federal Government and the States as well as the German Research Foundation (DFG) under Grant No. 495926269 is gratefully acknowledged. 





\bibliography{sn-bibliography}

\end{document}